\documentclass[a4paper,fleqn,usenatbib]{mnras}
\usepackage[T1]{fontenc}
\usepackage{ae,aecompl}

\usepackage{graphicx}	% Including figure files
\usepackage{amsmath}	% Advanced maths commands
\usepackage{amssymb}	% Extra maths symbols
\usepackage{bm}
\usepackage{subfig}
\usepackage{times}
\usepackage{color}
\usepackage{soul}
\newcommand{\bhac}{\texttt{BHAC}~}

\newcommand{\eg}{e.g.,~}
\newcommand{\ie}{i.e.,~}

\title[Plasmoid formation in GRMHD simulations]{Plasmoid formation in
  global GRMHD simulations and AGN flares}

\author[A. Nathanail et al.]{Antonios Nathanail$^{1}$, \thanks{E-mail:
    nathanail@itp.uni-frankfurt.de} Christian M. Fromm$^{1,2}$, Oliver
  Porth $^{3}$, Hector Olivares$^{1}$, \newauthor Ziri Younsi$^{4}$,
  Yosuke Mizuno$^{1}$ and Luciano Rezzolla$^{1,5}$ \\
$^{1}$Institut f\"ur Theoretische Physik, Goethe Universit\"at Frankfurt,
Max-von-Laue-Str.1, 60438 Frankfurt am Main, Germany \\
$^{2}$Max-Planck-Institut f\"ur Radioastronomie, Auf dem H\"ugel 69,
D-53121 Bonn, Germany \\
$^{3}$Astronomical Institute Anton Pannekoek, Universeit van Amsterdam, Science Park 904, 1098 XH, Amsterdam, The Netherlands \\
$^{4}$Mullard Space Science Laboratory, University College London,
Holmbury St. Mary, Dorking, Surrey, RH5 6NT, UK\\
$^{5}$School of Mathematics, Trinity College, Dublin 2, Ireland
}
\begin{document}
\label{firstpage}
\pagerange{\pageref{firstpage}--\pageref{lastpage}}
\maketitle

\begin{abstract}
One of the main dissipation processes acting on all scales in
relativistic jets is thought to be governed by magnetic
reconnection. Such dissipation processes have been studied in idealized
environments, such as reconnection layers, which evolve in merging
islands and lead to the production of ``plasmoids'', ultimately resulting
in efficient particle acceleration. In accretion flows onto black holes,
reconnection layers can be developed and destroyed rapidly during the
turbulent evolution of the flow. We present a series of two-dimensional
general-relativistic magnetohydrodynamic simulations of tori accreting
onto rotating black holes focusing our attention on the formation and
evolution of current sheets. Initially, the tori are endowed with a
poloidal magnetic field having a multi-loop structure along the radial
direction and with an alternating polarity. During reconnection
processes, plasmoids and plasmoid chains are developed leading to a
flaring activity and hence to a variable electromagnetic luminosity. We
describe the methods developed to track automatically the plasmoids that
are generated and ejected during the simulation, contrasting the
behaviour of multi-loop initial data with that encountered in typical
simulations of accreting black holes having initial dipolar field
composed of one loop only. Finally, we discuss the implications that our
results have on the variability to be expected in accreting supermassive
black holes.
\end{abstract}

%\keywords{gravitational waves, gamma-ray bursts}
%

%-----------------------------------------------------------------------
\section{Introduction} 
\label{sec:intro}
%-----------------------------------------------------------------------
%

Relativistic jets are observed in high-energy sources, from gamma-ray
bursts (GRB) to active galactic nuclei (AGNs). They can be launched from
magnetic processes around black holes
\citep{Blandford1977,Blandford:1982di} and the numerical simulation of
these processes has now reached significant maturity (see, \eg
\citealt{Rezzolla:2011,Akiyama2019_L5}). The magnetic-field topology
plays a central role in determining the structure of the outflows from
these systems and is fundamental for energy dissipation through various
magnetohydrodynamic (MHD) instabilities and magnetic reconnection.

Accretion in magnetized disks is believed to be driven by the
magneto-rotational instability (MRI) \citep{Balbus1991}, where the
advection of plasma and magnetic fields provides the main ingredients to
launch magnetized winds and jets. Magnetically dominated jets accelerate
efficiently the bulk of the plasma while keeping much of the energy
stored into the field itself \citep{Komissarov:2009, Lyubarsky2009,
  Tchekhovskoy2009}. Under these conditions, magnetic reconnection
represents an efficient way to dissipate some of this magnetic energy and
it has been proposed to explain GRB and AGN emission \citep{diMatteo1998,
  Zhang2011, Giannios2013, Dionysopoulou2015}

Large-scale magnetized jets that are produced from accretion flows around
black holes tend to have a variable electromagnetic power. This is due to
the intrinsically turbulent nature of the accretion process that, in
turn, produces and advects onto the black hole magnetic-field loops of
different polarity, whose interaction provides the sites for formation of
current sheets. These magnetic loops can emerge in the surface of the
disk due to buoyancy like the Rayleigh-Taylor instability
\citep{Parker1966}, and can change the topology of the magnetic field
throughout the jet. Furthermore, because of their turbulent genesis,
these will not respect any symmetry across the equatorial plane and may
therefore lead to differences in the wind properties above and below the
equatorial plane \citep{Kadowaki2018}.

Previous studies have established the importance of the magnetic-field
configuration for the production of a steady outflow and more
specifically its power. The nested-loop poloidal magnetic-field
structure, which is the one normally adopted as initial data to model
magnetized accretion onto black holes \citep{Gammie:2003qi,
  DeVilliers2005, McKinney2006}, has been shown to produce relativistic
jets. Depending on the initial strength of the magnetic field, the
efficiency of the energy extraction from the black hole can go beyond
$100\%$ efficiency, thus producing very powerful outflows in the case of
magnetically arrested disks (MAD, \citealt{Igumenshchev2003, Narayan2003,
  Tchekhovskoy2011}).

In addition to the nested-loops, different magnetic-field configurations
have also been studied, leading to a picture in which jet launching is
especially sensitive to the initial magnetic-field geometry
\citep{Beckwith2008,Beckwith2009}. More specifically, the outflows
resulting from the accretion can vary depending on the the initial
magnetic-field configuration, going from jets that are weak and mostly
turbulent, to powerful and collimated ones \citep{Beckwith2008,
  McKinney2009, McKinney2012, Narayan2012, Liska2018b, Akiyama2019_L5}. A
particularly interesting configuration that has been studied recently is
one in which loops of alternating polarity are periodically advected to
the black hole \citep{Parfrey2015}. Such configurations with a region of
alternating polarity can be also produced naturally in accretion flows
around black holes \citep{Contopoulos1998, Contopoulos2015,
  Contopoulos2018}, and lead rather naturally to the development of
regions of alternating magnetic-field polarity.

Recent observations of rapid variability of X-ray/gamma-ray flares in
blazars with timescales from several minutes to a few hours pose severe
constraints on the particle acceleration timescale and the size of the
emission region (\eg \citealt{Aharonian2007,Albert2007}). From
observations, fast variable flares may come from a small region with a
size of the order of a few Schwarzschild radii, launching fast moving
``needles'' within a slower jet or from jet within a jet
\citep{Levinson2007,Begelman2008, Ghisellini2008,Giannios2009}. Flares
also require very rapid particle acceleration.

Throughout the jet, instabilities -- and the associated turbulence -- can
can provide rather naturally the sites for the generation of current
sheets and hence for the occurrence of magnetic reconnection. These
configurations of current sheets structured by large sheets of
alternating magnetic-field polarity have been studied and found to be
very efficient in particle acceleration (see, \eg \citealt{Kagan2015} for
a review). Quite generically the initial configuration fragments at the
current sheet layer and produces chains of magnetic ``plasmoids'' (or
``magnetic islands'') \citep{Loureiro2007, Uzdensky2010, Fermo2010,
  Huang2012, Loureiro2012, Takamoto2013}. We recall that plasmoids are
quasi-spherical regions that contain relativistic particles and have a
large magnetization, that is a large ratio between the magnetic and
rest-mass energies. Magnetic reconnection and plasmoid formation has
been extensively studied in Particle-in-Cell (PIC) simulations, which
have confirmed that magnetic reconnection in relativistic regimes to be
very efficient in accelerating charged particles \citep{Sironi2014,
  Guo2014, Guo2015, Werner2016, Sironi2016, Li2017PIC, Kagan2018,
  Petropoulou2019}. Plasmoids produced from magnetic reconnection events
can be studied systematically in a statistical manner
\citep{Petropoulou2018} and have been introduced to explain flaring
activity and AGN emission \citep{Petropoulou2016, Zhang2018a,
  Christie2019}.

\begin{figure*}
  \begin{center}
    \includegraphics[width=0.99\textwidth]{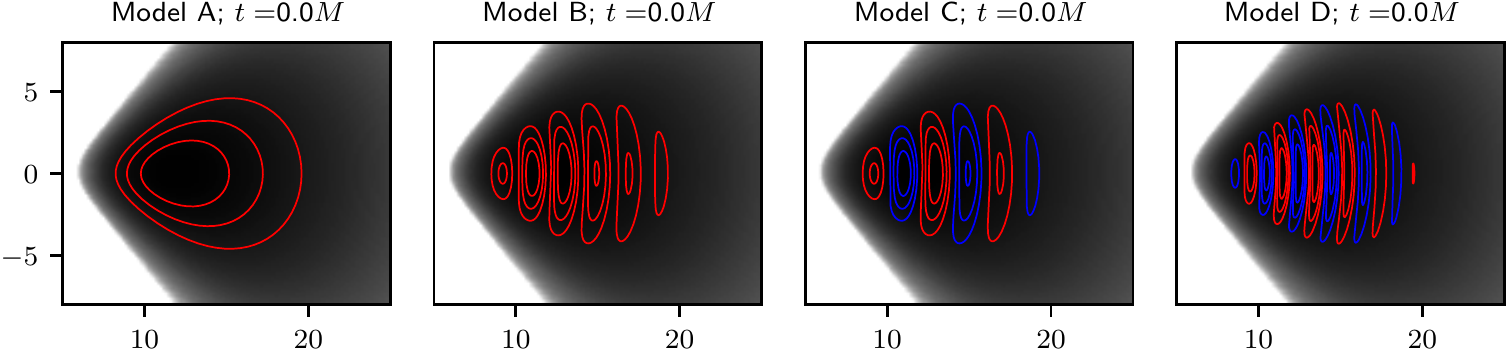}
  \end{center}
  \caption{Initial magnetic-field configuration for the four models.
    from left to right is \texttt{\small{Model A}}, \texttt{\small{Model
        B}}, \texttt{\small{Model C}} and \texttt{\small{Model D}}. The
    different colors of the magnetic loops indicate their direction,
    where red is clockwise and blue counterclockwise. In the background
    the density of the torus is shown in black.}
    \label{fig:init}
\end{figure*}

In view of the potential importance of plasmoid formation and evolution
in the dynamics and energetics of relativistic jets, we have here carried
out the first steps to asses and measure such properties in rather
realistic general-relativistic conditions of accretion flows around black
hole. The main goal of this work is, therefore, is to build and explore
those magnetic-field geometries that are more likely to lead to the
formation of current sheets and therefore to magnetic reconnection.

To this scope, we have performed a series of two-dimensional
general-relativistic ideal-MHD simulations exploring initial magnetic
fields having different coherence length and topology, and assessing the
impact that these initial conditions have on the production rate of
plasmoids in each case. More specifically, we initialize the accretion
torus with a varying number of poloidal magnetic loops of alternating
polarity and contrast the results of the corresponding simulations with
those obtained with the nested-loop setup typically used in the
literature. Special attention is paid to the formation of current sheets,
to the occurrence of magnetic reconnection, and to the consequent
production of plasmoids and plasmoid chains. Overall, we find that
reconnection layers are rapidly developed and destroyed in the vicinity
of the black hole. In such reconnection layers, plasmoids are generated
and, in some cases, accelerated to large energies, thus becoming
candidates to explain the flaring activity in AGNs.
 
This paper is organized as follows: in Section \ref{sec:num} we present
the overview of the simulations, the numerical setup \ref{sec:setup}, a
brief comparison of the initial models \ref{sec:comp} and the specifics
of the models where reconnection occurs \ref{sec:lay}. Section
\ref{sec:plasmoids} discussed in more detail the production of plasmoids
through the reconnection layers and their evolution. Finally, we
summarize and present a discussion about the results in Section
\ref{sec:con}.

%-----------------------------------------------------------------------
\section{Numerical details and model comparison} 
\label{sec:num}
%-----------------------------------------------------------------------

%
\subsection{Numerical setup} 
\label{sec:setup}
%
%\begin{figure*}
%  \begin{center}
%    \includegraphics[width=0.95\textwidth]{test_initial_loops-crop}
%  \end{center}
%  \caption{Initial magnetic field configuration for the four models. 
%  from left to right is \texttt{\small{Model A}}, \texttt{\small{Model B}},
%  \texttt{\small{Model C}} and \texttt{\small{Model D}}. The different 
%  colors of the magnetic loops indicate their direction, where red 
%  is clockwise and blue counterclockwise. In the background the density of the 
%  torus is shown in black.}
%    \label{fig:init}
%\end{figure*}

For our simulations, we employ \bhac \citep{Porth2017}, which solves the
equations of general-relativistic ideal MHD using second-order
high-resolution shock-capturing finite-volume methods. The code has been
employed in a number of investigations, \eg \citep{Mizuno2018,
  Nathanail2018c}, and has been carefully tested and compared with codes
with similar capabilities \cite{Porth2019}. \bhac solves the
general-relativistic MHD equations
\begin{align}
\nabla_{\mu}(\rho u^{\mu}) &= 0\,, \\
\nabla_{\mu} T^{\mu\nu} &= 0\,,   \\
\nabla_{\mu} \,^{\ast}F^{\mu\nu} &= 0\,, 
%% \nabla_{\mu} \,F^{\mu\nu} &= 4\pi J^{\nu} \,,
\label{gen_eqs}
\end{align}
where $\rho$ is the rest-mass density, $u^{\mu}$ the fluid $4$-velocity,
$T^{\mu\nu}$ the energy momentum tensor and $^{\ast}F^{\mu\nu}$ is the
dual of the Faraday tensor. Our simulations are performed in two spatial
dimensions. The code makes uses of fully adaptive mesh-refinement (AMR)
techniques and of the constrained-transport method \citep{DelZanna2007}
to preserve a divergence-free magnetic field \citep{Olivares2019}.

As initial data we consider an axisymmetric equilibrium torus with
constant specific angular momentum~\citep{Fishbone76} $\ell=4.28$ around
a Kerr black hole with a dimensionless spin of $a:=J/M^2=0.93$, where $J$
and $M$ are the angular momentum and mass, respectively. The inner radius
of the torus is set to be $r_{\rm in}=6\, r_g$ and the outer radius
$r_{\rm out}=12\, r_g$, where $r_g:= M$ is the gravitational radius (we
use units in which $G=c=1$).

As mentioned above, the initial magnetic field is buried in the torus and
purely poloidal, but chosen so as to have four different topologies
consisting of a series of nested loops with varying polarity along the
radial direction. This is achieved by making use of a vector potential
with the form
%
%% \begin{align}
%% A_{\phi} \propto &\max(\rho/\rho_{\rm max}-0.2,0) \nonumber \\
%% &\times \cos((N -1)\theta) \sin(2\pi \, (r-r_{\rm in} )/\lambda_r) \nonumber \\
%% %
%% \propto &\mathcal{A}_0\times\mathcal{B}_0         \,, \label{eq:aphi}
%% \end{align}
%% %

%
\begin{align}
&A_{\phi}\propto \mathcal{A}\times\mathcal{B}  \\
&\mathcal{A} = \max(\rho/\rho_{\rm max}-0.2,0) \nonumber \\
&\mathcal{B} = \cos((N -1)\theta) \sin(2\pi \, 
(r-r_{\rm in} )/\lambda_r) \nonumber          \,, 
\end{align}
%
%% \an{Check notation, keep one eq. above}
where $\rho_{\rm max}$ is the maximum rest-mass density in the torus and
the parameters $N\geq 1$ and $\lambda_r$ set the number and the
characteristic lengthscale (and hence the polarity) of the poloidal loops
inside the torus respectively and thus are varied to produce the
desirable initial magnetic-field topologies.

Overall, we have considered four different models whose initial
magnetic-field structure is reported in Fig. \ref{fig:init}. In most of
the runs we use a logarithmic grid in the radial direction so that the
domain extends to $2500\, r_g$. The various parameters used are
summarized in Table \ref{table}, where the last three columns refer to
the resolution of the simulations, which is different for the various
setups because of the varying characteristic lengthscales in the. Most of
the figures discussed hereafter will refer to simulations performed with
the highest resolution: \ie at the base resolution for
\texttt{\small{B}},~ at $4\times$ the base
resolution for \texttt{\small{Models A}}, \texttt{\small{C}}  
and \texttt{\small{D}}.

\begin{table*}
  \centering
   \caption{Initial parameters for the various models considered. The
     first three columns are the parameters that set the initial magnetic
     field configuration. The next column corresponds to the base
     resolution for every model and in the remaining columns the
     respective models that were run in higher resolution for
     comparison.}
  \begin{tabular}{l|c|c|l|c|c|c|c|c|}
    \hline
    \hline
     model  & $N$  &
     $\lambda_r$  & $\hskip 0.5cm A_{\phi}$ &  $N_r \times  N_{\theta}
     \times  N_{\phi} $ & $N_r \times  N_{\theta}
     \times  N_{\phi} $ & $N_r \times  N_{\theta}
     \times  N_{\phi} $ & $N_r \times  N_{\theta}
     \times  N_{\phi} $  \\
     &  & & & (base res.) & (2 $\times$ base res.)  & (4 $\times$ base
     res.) & (6 $\times$ base res.)  \\
 %    & $[M_\odot]$  & $[\mathrm{km}]$  &
 %    $[\mathrm{Hz}]$  & $-$    \\
    \hline
  \texttt{\small{Model A}} & $1$ & $-$ & $A_{\phi}=\mathcal{A}$  & $ 1024  \times 
   512  \times  1$ & $ 2048  \times  1024  \times  1$& $
  4096  \times  2048  \times  1$&\\
  \texttt{\small{Model B}} & $3$ & $4$ & $A_{\phi}=\mathcal{A}\times|\mathcal{B}|$  & $ 1024  
  \times  512  \times  1$ & $-$ & $-$ &$-$\\
  \texttt{\small{Model C}} & $3$ & $4$ & $A_{\phi}=\mathcal{A}\times\mathcal{B}$  & $ 1024  \times 
   512  \times  1$ & $ 2048  \times 
   1024  \times  1$& $ 4096  \times 
   2048  \times  1$ & $-$ \\
  \texttt{\small{Model D}} & $3$ & $2$ & $A_{\phi}=\mathcal{A}\times\mathcal{B}$  & $ 1024  \times 
   512  \times  1$ & $ 2048  \times 
   1024  \times  1$& $ 4096  \times 
   2048  \times  1$& $ 6144  \times 
   3072  \times  1$\\
%  \texttt{\small{Model C}}$_{3D}$ & $3$ & $4$ & $-$ & $100\, r_g$ & $ 1024 \, \times 
%  \, 512 \, \times \, 96 $ & $-$ & $-$ \\
%	  \texttt{\small{Model D}}$_{3D}$ & $3$ & $2$ & $-$ & $2500\, r_g$ & $ 512 \, \times 
%  \, 256 \, \times \, 256 $ & $-$ & $-$ \\
   \hline
    \hline
  \end{tabular}
  \label{table}
\end{table*}

The first model, \texttt{\small{Model A}}, includes the typical nested
loop magnetic-field topology. \texttt{\small{Model B}} consists of a
multi-loop structure where all loops have the same polarity. The last two
models have a multi-loop structure where each loop has an alternate
polarity. The loops of \texttt{\small{Model B}} and \texttt{\small{Model
    C}} have a similar width, whereas in \texttt{\small{Model D}} the
loops have a smaller size. For the last two models, several resolutions
where used to check the impact on the activation and saturation of the
MRI, these are discussed in the Appendix \ref{appen}. Furthermore, for
the two models with alternate polarity loops, \texttt{\small{Model C}}
and \texttt{\small{Model D}}, we run a set of $3D$ simulations. The
simulation runs are evolved up to $t=5 \, \times \, 10^3\,M$. Such a
timescale is sufficiently long to capture all the important and
distinctive features of each simulation and is sufficiently short that
the we need not to be concerned about the decaying poloidal magnetic
field, which takes place in axisymmetry as a consequence of the decay of
turbulence \citep{Cowling33, Sadowski2015}.

\subsection{Model comparison} 
\label{sec:comp}

 \begin{figure}
  \begin{center}
    \includegraphics[width=1.05\columnwidth]{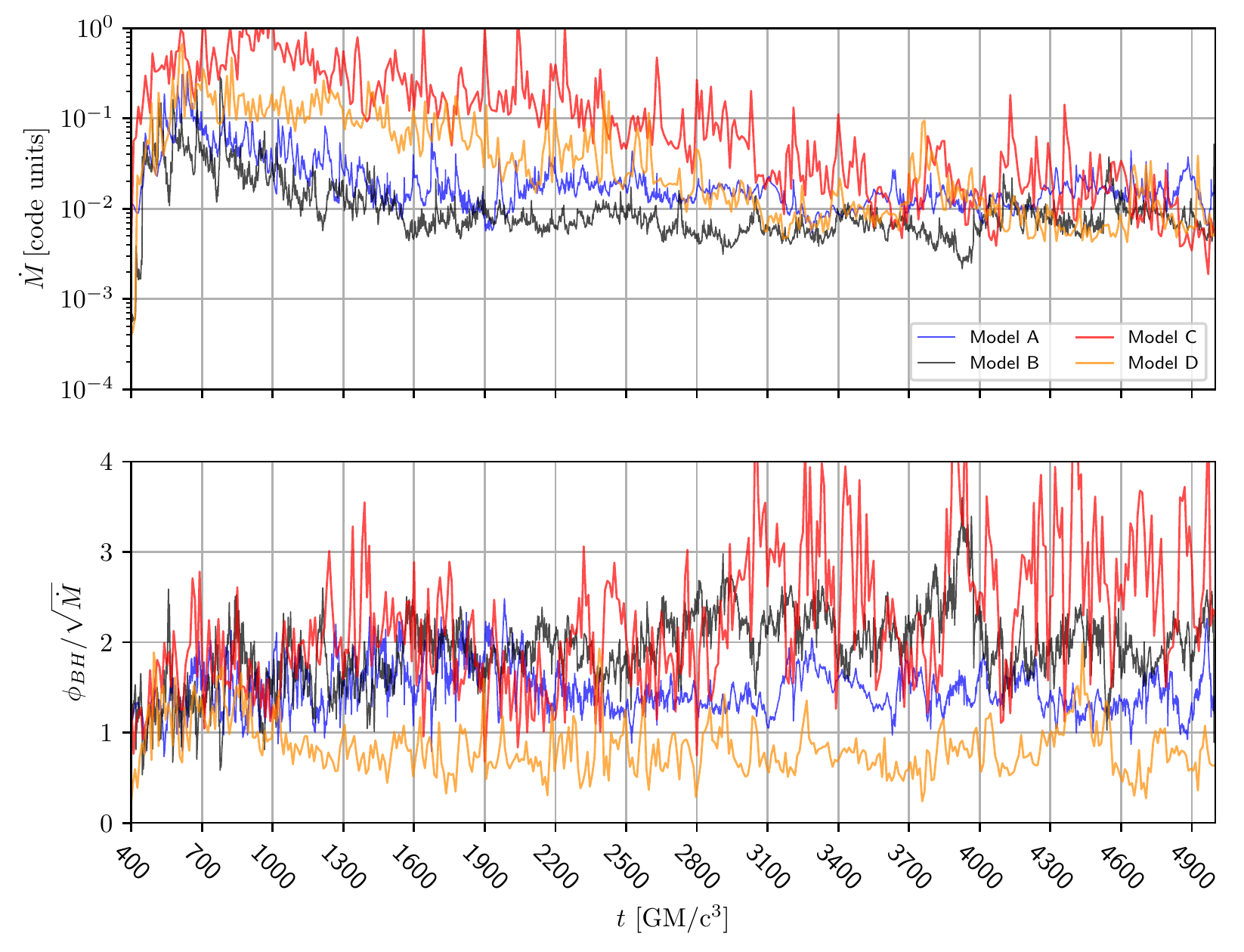}
  \end{center}
  \caption{Upper panel: the rate of mass accreted through the black-hole
    horizon, Lower panel: the magnetic flux accumulated on the black-hole
    horizon. For models \texttt{\small{Models A, B, C }} and
    \texttt{\small{D}}, reporting the high-resolution run for each model
    (\ie at the base resolution for \texttt{\small{Models A}} and
    \texttt{\small{B}},~ at $4\times$ the base resolution for
    \texttt{\small{Model C}}, and at $6\times$ for \texttt{\small{Model
        D}}).}
    \label{fig:mdot}
\end{figure}
\begin{figure}
  \begin{center}
    \includegraphics[width=1.05\columnwidth]{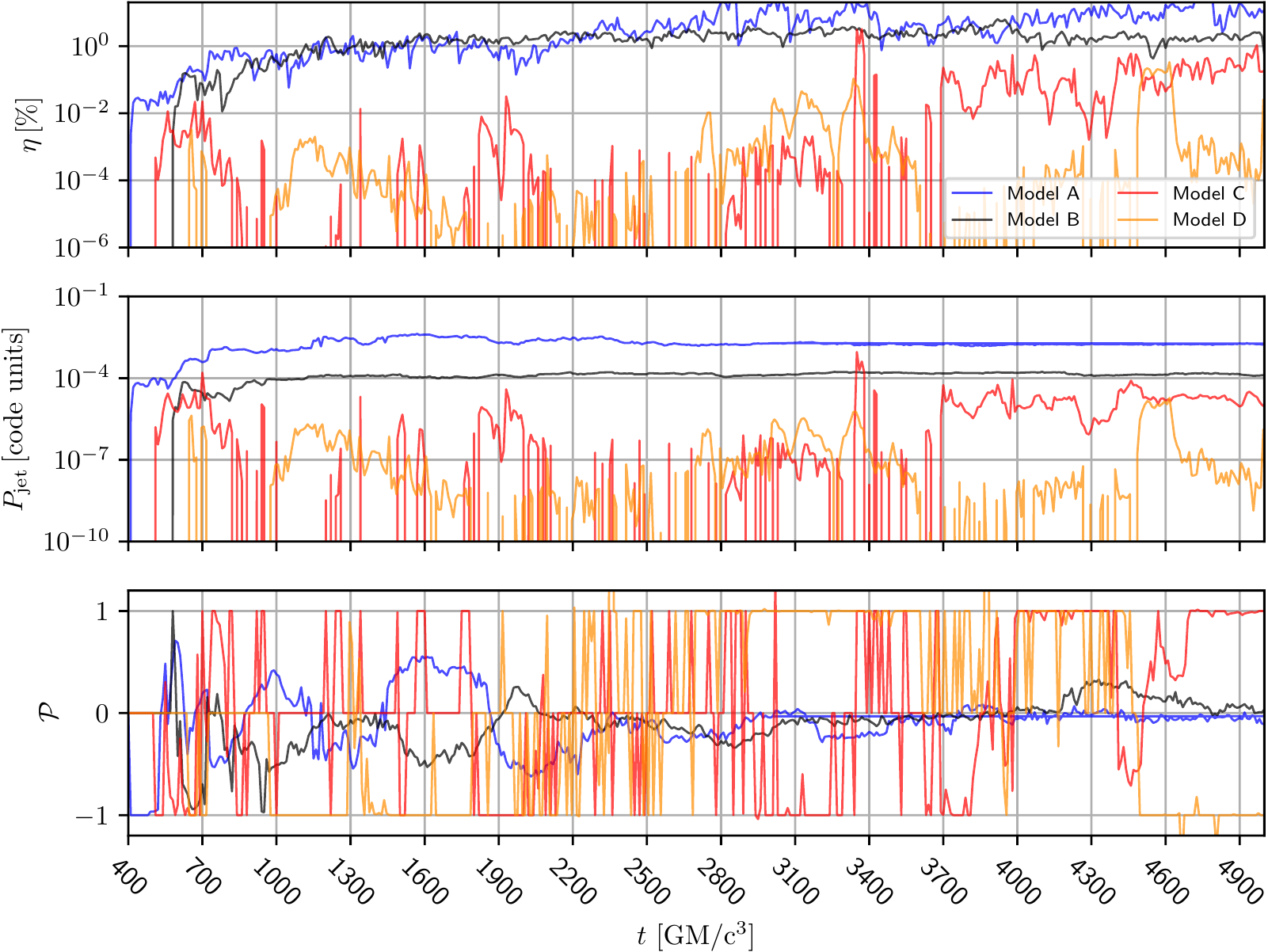}
  \end{center}
  \caption{Upper panel: the efficiency of the outflow as defined in
    Eq. \eqref{eff}, Middle panel: the power of the jet as defined in
    Eq. \eqref{pjet}, Lower panel: the difference in the power, of the
    upper and the lower jet as defined in Eq. \eqref{pjetns}.For models
    \texttt{\small{Model A, B, C }} and \texttt{\small{D}}, referring to
    the highest-resolution run for each model.}
    \label{fig:eta_plasm}
\end{figure}

Since our focus here is to determine and highlight those features in the
plasma dynamics that emerge when considering different initial
magnetic-field topologies, we first discuss the main differences among
the four models. A particularly useful quantity in this respect are the
amount of rest-mass and magnetic field accreted onto the black hole,
namely, the mass-accretion rate and the accreted magnetic flux. The
former is measured as
\begin{align}
 \dot{M}:=\int_0^{2\pi}\int^\pi_0\rho u^r\sqrt{-g}\,d\theta d\phi\,,
\label{mdot}
\end{align}
and is shown for all models in the upper panel of Fig.
\ref{fig:mdot}. Note that after a time of $\sim 3000\, M$, the accretion
rate in all models relaxes to a roughly constant value. In
\texttt{\small{Model C}} and \texttt{\small{D}} the mass-accretion rate
remains roughly constant till $12000\, M$, in the high-resolution runs,
thus reflecting an essentially stationary turbulent state in the torus.

Similarly, we define magnetic flux accreted across the event horizon as
\begin{align}
  \phi_{\rm BH}:=\frac{1}{2}\int_0^{2\pi}\int^\pi_0 |B^r|\sqrt{-g}d\theta d\phi\,,
\label{phiBH}
\end{align}
and show its normalized value $\phi_{\rm BH}/\sqrt{\dot{M}}$ in the lower
panel of Fig. \ref{fig:mdot} for all of the models considered. Note that
for \texttt{\small{Model A}} the normalized magnetic flux is roughly
constant in time and it is far below the saturation limit $\phi_{\rm
  BH}=\phi_{\rm max}\approx 15$ that is normally taken for a MAD
configuration (within the units adopted here
\cite{Tchekhovskoy2011}). This net magnetic flux is a prerequisite for
energy extraction from the black hole \citep{Blandford1977}.
%,Nathanail2014} In the other runs the magnetic flux exhibits some variability.
What the simulations reveal is that magnetic flux of one polarity is
brought towards the horizon of the black hole but also that magnetic flux
of the opposite polarity reaches the vicinity of the horizon, thus
annihilating the previous one and reducing the overall flux across the
horizon. Fluctuations in this overall behaviour are stronger in the two
cases where the initial loops have larger width, \ie
\texttt{\small{Models B}} and \texttt{\small{C}}. In these cases, the
evolution of the magnetic flux shows both a short dynamical timescale --
as magnetic flux is brought to the horizon -- but also a longer timescale
reflecting the amount of time needed for the whole loop to be partly
annihilated and as a result reduce the magnetic flux on the
horizon. Overall, the magnetic flux in \texttt{\small{Models B}} and
\texttt{\small{C}} is at all times higher than in the other two cases. It
is important to note here that even if the magnetic flux in
\texttt{\small{Model C}} is three or four times larger than
\texttt{\small{Model A}}, a magnetic funnel is never produced due to the
continuous magnetic flux annihilation.

All models considered produce outflows that give off energy at
infinity. This can be measured through the energy flux that passes
through a 2-sphere placed at $50 \, r_g$
\begin{align}
  P_{\rm jet}:=\int_0^{2\pi}\int^\pi_0 (-T^r_t -\rho u^r)\sqrt{-g}d\theta
  d\phi\,,
\label{pjet}
\end{align}
where the integrand in \eqref{pjet} is set to zero if everywhere on the
integrating surface $\sigma \leq 1$. The efficiency of the jet is then
defined as
\begin{align}
  \eta :=P_{\rm jet}/\dot{M} \,,
\label{eff}
\end{align}
and can become larger than unity. Lastly, in order to measure the
contribution, and difference in the power from the upper and the lower
jet we introduce the jet asymmetry
\begin{align}
  \mathcal{P}:= \frac{P_{\rm jet,u}- P_{\rm jet,l}}{(P_{\rm jet,u} + P_{\rm jet,l})}\,,
\label{pjetns}
\end{align}
where $P_{\rm jet,u}$ and $P_{\rm jet,l}$ are the powers of the upper and
lower jets, respectively. Clearly, $\mathcal{P} \simeq 1$ when the upper
jet dominates, $\mathcal{P} \simeq -1$ for power releases dominated by
the lower jet, while $\mathcal{P} \simeq 0$ refers to a symmetric jet
emission. Note that we set $\mathcal{P}=0$ when the power of both the
upper and the lower jets are zero because nowhere on the integrating
surface is the condition $\sigma > 1$ satisfied.
 
\begin{figure*}
  \begin{center}
    \includegraphics[width=0.95\textwidth]{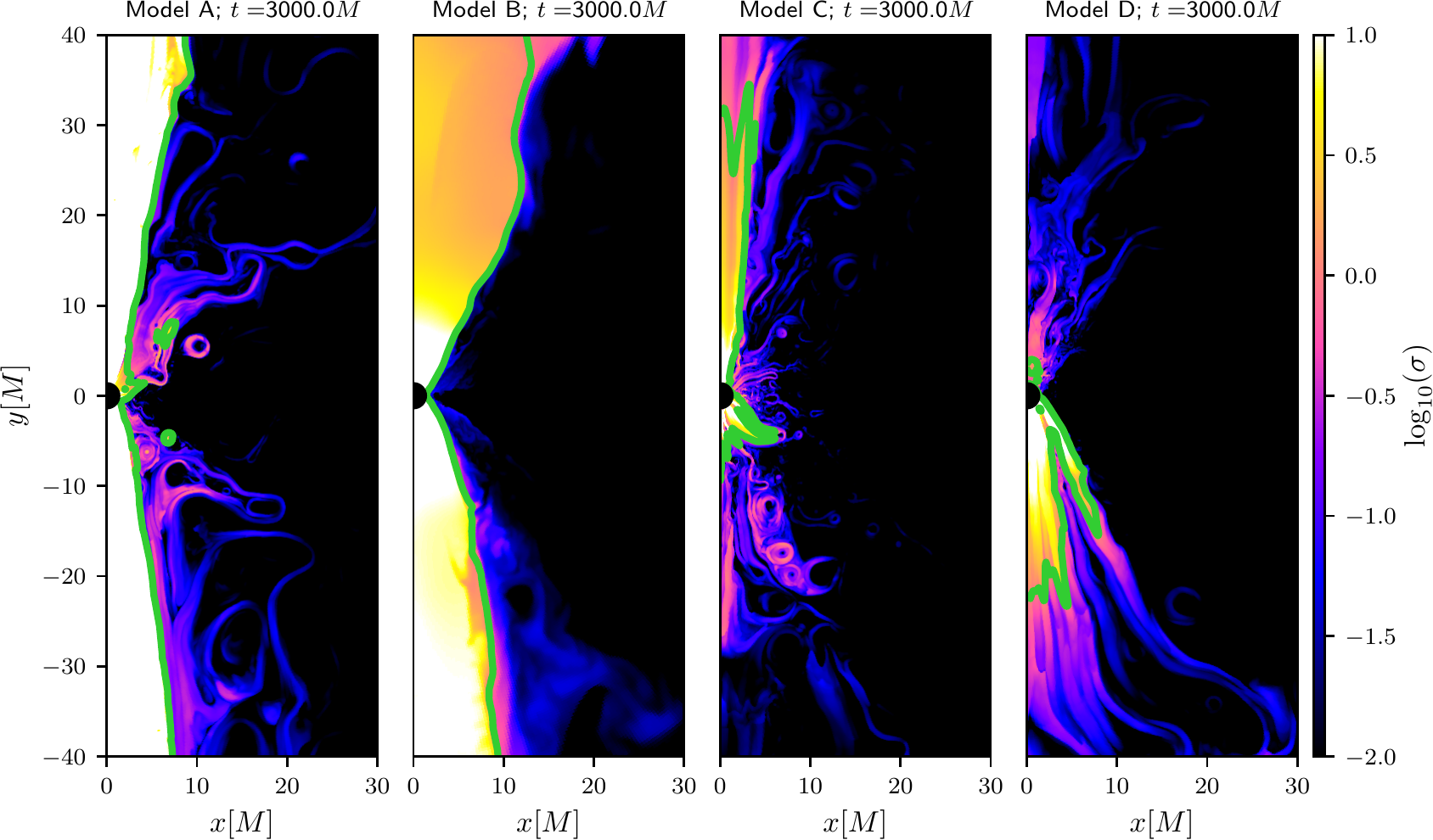}
    \includegraphics[width=0.95\textwidth]{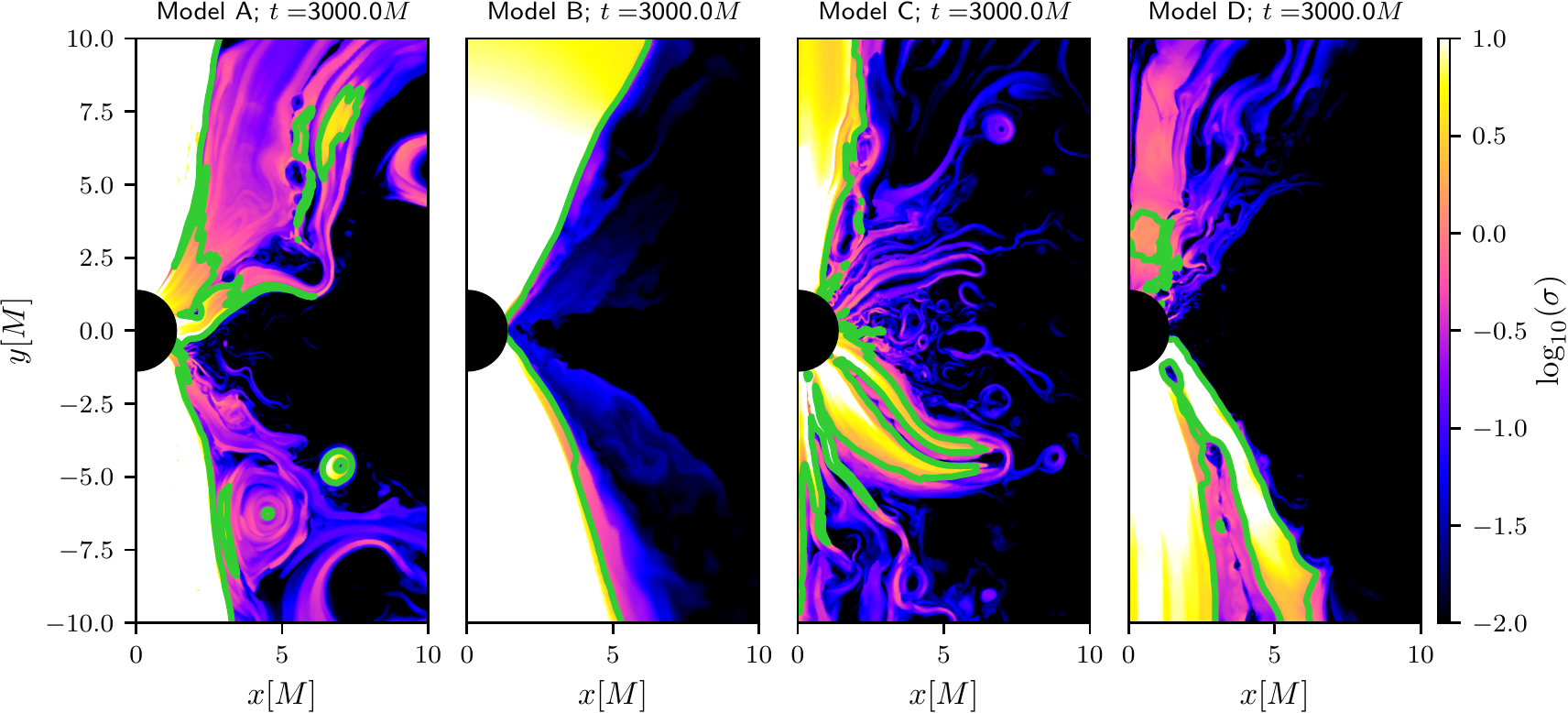} 
  \end{center}
  \caption{The magnetization parameter $\sigma$ for the four models at
    time $t=3000 \, M$. The lower panels show a magnified view near the
    black hole. The green line shows the contour of $\sigma=1$.  The
    resolutions are: base resolution for \texttt{\small{Model B}},
    $4\times$ base resolution for \texttt{\small{Models A}},
    \texttt{\small{C}} and \texttt{\small{D}}.}
    \label{fig:sigma}
\end{figure*}

\begin{figure*}
  \begin{center}
    \includegraphics[width=0.95\textwidth]{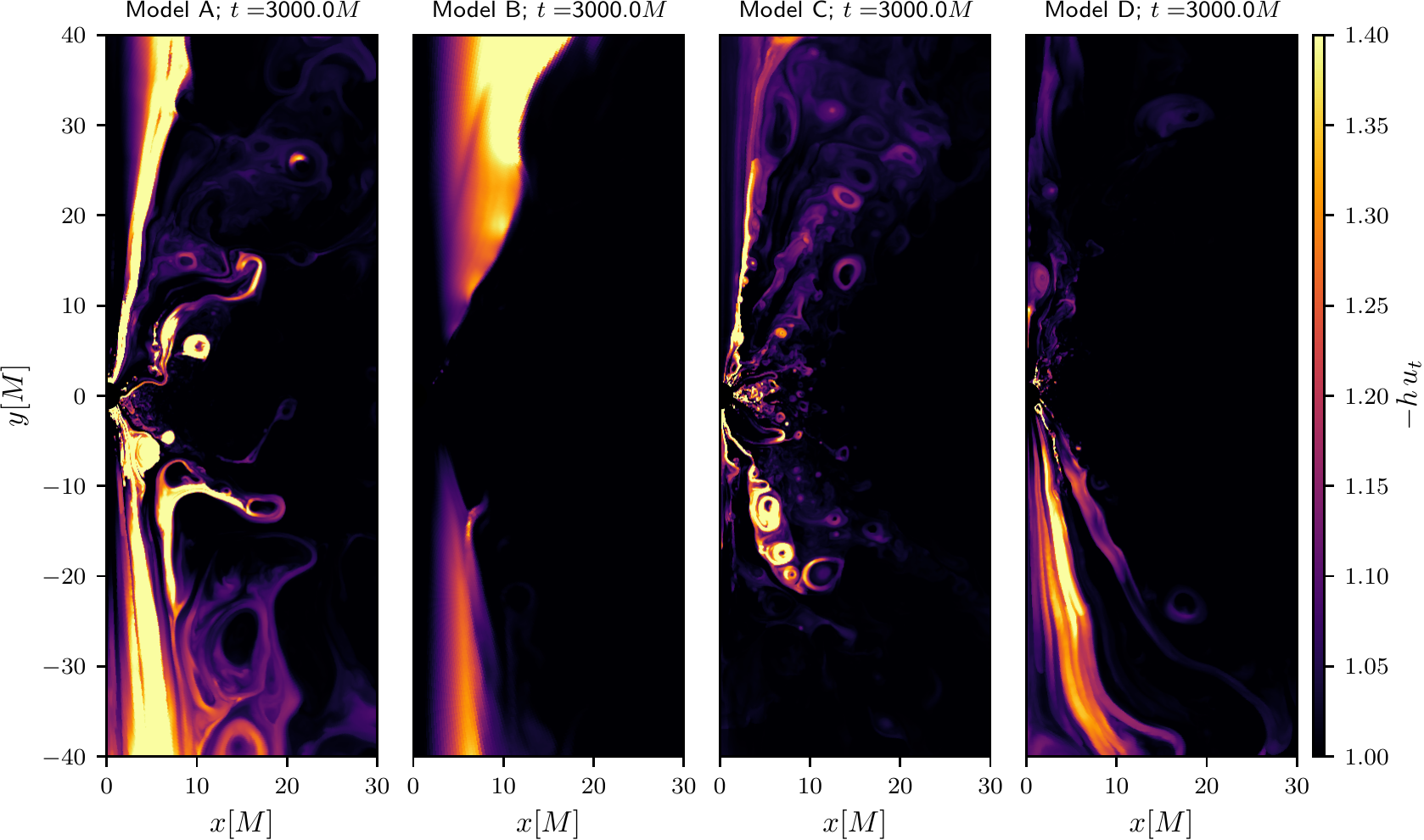}
  \end{center}
  \caption{Bernoulli constant $-hu_t$ at time $t=3000 \, M$, any
    feature/fluid element which is not black in the figure is
    unbound/outgoing. The two models on the left have a well-defined and
    uniform regions where matter is unbound, whereas the models on the
    right have smaller unbound regions mostly localised in plasmoids. The
    resolutions are: base resolution for \texttt{\small{Model B}},
    $4\times$ base resolution for \texttt{\small{Models A}},
    \texttt{\small{C}} and \texttt{\small{D}}.}
    \label{fig:hut}
\end{figure*}

All of these quantities are shown in the three panels of
Fig. \ref{fig:eta_plasm} for the various models considered. In
particular, it is possible to note that \texttt{\small{Models A}} and
\texttt{\small{B}} produce steady jet structures where the efficiency of
the jet and the power of the jet are roughly constant in time (see upper
and middle panels of Fig. \ref{fig:eta_plasm}). Furthermore, the
difference of the upper and lower jet is at all times close to zero
(lower panel of Fig. \ref{fig:eta_plasm}), which means that both jets
contribute equally to the overall power. \texttt{\small{Models A}} and
\texttt{\small{B}} are quite similar in jet power and efficiency. This is
because the small-scale field between the loops of \texttt{\small{Model
    B}} reconnect and yield a topology similar to \texttt{\small{Model
    A}}. The magnetic field in \texttt{\small{Models C}} and
\texttt{\small{D}}, on the other hand, do not reconnect so efficiently
and the long-term dynamics is governed by the small-scale seed fields.
Furthermore, for \texttt{\small{Models C}} and \texttt{\small{D}} there
is a strong variability in the power of the outflows and as we will see
below they do not even form steady structures. It is important to note
that even if they do not produce jets, in the usual form, they exhibit an
intense flaring activity. It is tempting to relate the timescale of the
flaring activity in \texttt{\small{Models C}} and \texttt{\small{D}} with
the characteristic lengthscale $\lambda_r$ of the initial magnetic
loops. In the case of \texttt{\small{Model C}}, the loops have a larger
coherence lengthscale which is then imprinted on the subsequent turbulent
flow and leads to typical times between flares to be longer (Fig.
\ref{fig:eta_plasm}). Similarly, for \texttt{\small{Model D}}, where the
loops are smaller the power is highly variable and changes in a smaller
timescale.

Using the mass estimate for Sgr~A$^*$ coming from \cite{Boehle2016}, this
flaring activity has a time span from minutes to tens of minutes, and
could therefore be correlate with the recent observations of the X-ray
activity of Sgr A$^*$ \citep{Haggard2019}. Some variability is seen in
well structured jets, however in the case of \texttt{\small{Models C}}
and \texttt{\small{D}} variability is very intense and the power during
the flare can vary up to two orders of magnitude in a time span of tens
of minutes for Sgr A$^*$ \citep{Do2019}.

In order to delimit and visualize the region representing the jet, we use
the magnetization parameter, $\sigma := B^2/\rho$, and define as jet the
low-density, strong magnetic-field region with $\sigma > 1$. In
Fig. \ref{fig:sigma} we plot $\sigma$ for all models in a logarithmic
colorcode and at time $t=3000,\,M$. Note that the jet is well structured
in \texttt{\small{Model A}} and that this structure is stationary
throughout the simulation. This is not particularly surprising and has
been shown to occur in numerous simulations over the last decade,
\citep[see, \eg][]{DeVilliers03b,McKinney2004,Porth2019}.
 
Similarly, in the case of \texttt{\small{Model B}}, where the initial
magnetic loops are of the same polarity, a similar magnetized region is
observed, which however has a reduced magnetization and a slightly
different shape. The magnetization is less than the previous model, with
$1<\sigma<10$, but clearly forms a jet structure. On the other hand, in
the other two case for \texttt{\small{Models C}} and \texttt{\small{D}},
no stationary magnetized structure is produced during the timescale of
the simulations. At all times, in fact, the funnel structure varies due
as reflected by the variability in the magnetic flux on the black-hole
horizon. Furthermore, the advection of magnetic loops of varying polarity
from the turbulent disc does no longer show a symmetry across the
equatorial plane. As a consequence, at some times a magnetized region can
be seen in the upper hemisphere, whereas at other times in the lower
hemisphere. Another important feature that can be identified in these
plots, is the formation of ``plasmoids'', whose properties we will
discuss in more detail below. magnetization.

\begin{figure*}
  \begin{center}
    \includegraphics[width=0.95\textwidth]{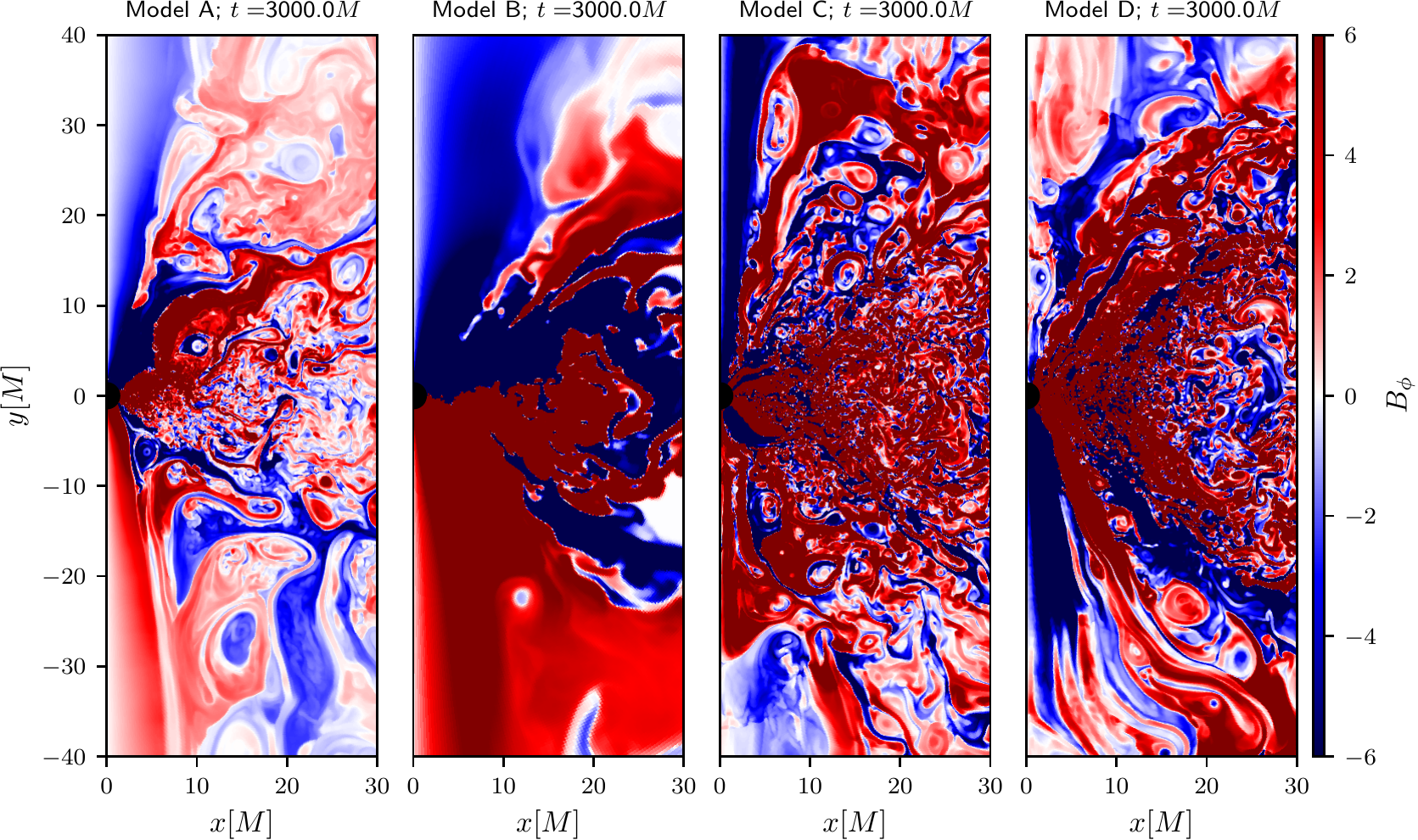}
  \end{center}
  \caption{The $B_{\phi}$ component of the magnetic field for the four
    models at time $t=3000 \, M$. The red and blue regions denote regions
    where the magnetic field has different polarity. The resolutions are:
    base resolution for \texttt{\small{Model B}}, $4\times$ base
    resolution for \texttt{\small{Models A}}, \texttt{\small{C}} and
    \texttt{\small{D}}. }
    \label{fig:Bphi}
\end{figure*}

\begin{figure*}
  \begin{center}
    \includegraphics[width=0.95\textwidth]{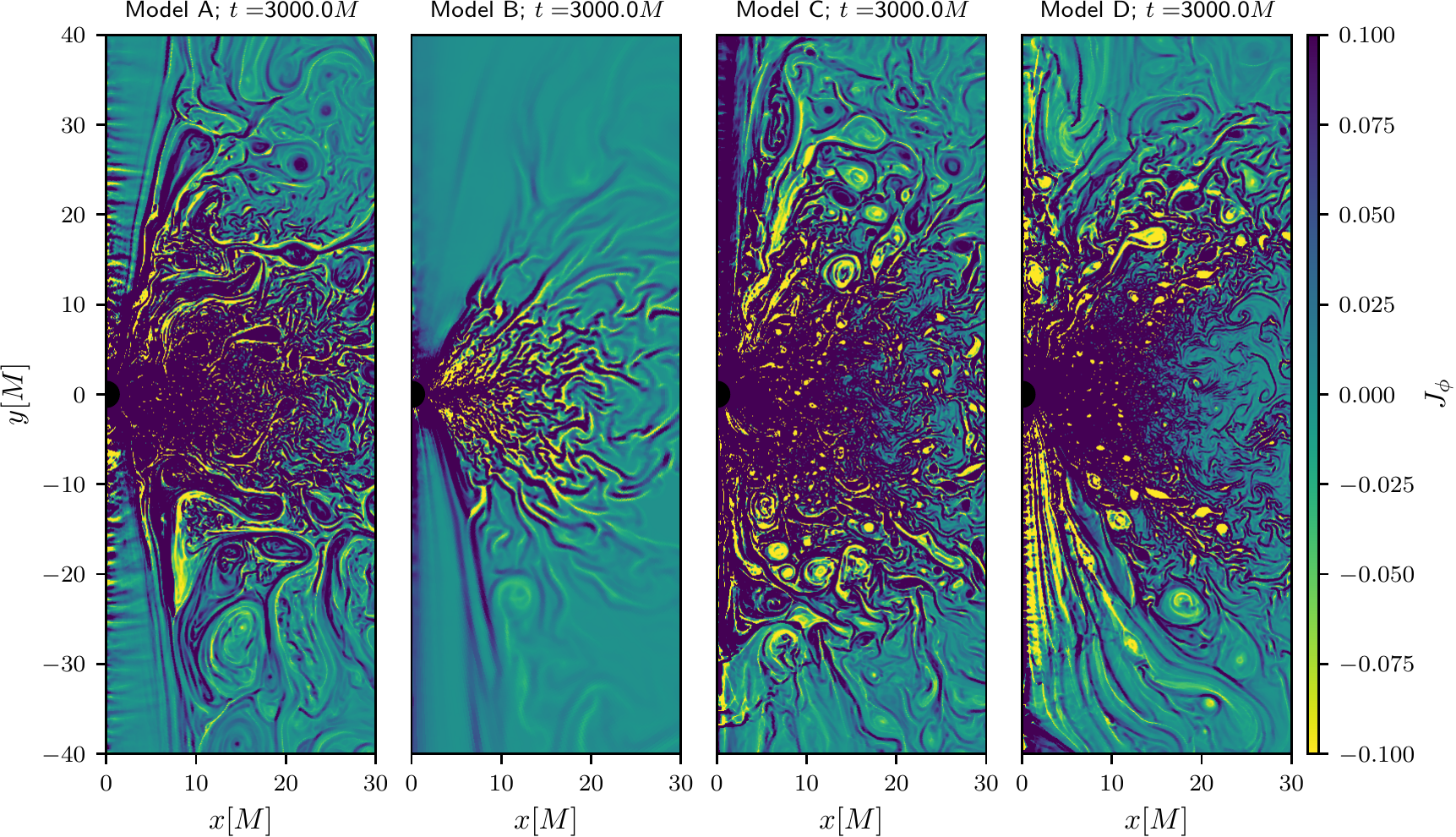}
  \end{center}
  \caption{The azimuthal component of the current $J_{\phi}$ for the four
    models at time $t=3000 \, M$. The resolutions are: base resolution
    for \texttt{\small{Model B}}, $4\times$ base resolution for
    \texttt{\small{Models A}}, \texttt{\small{C}} and
    \texttt{\small{D}}.}
    \label{fig:jphi}
\end{figure*}

\begin{figure*}
  \begin{center}
    \includegraphics[width=0.8\textwidth,height=0.8\textheight]{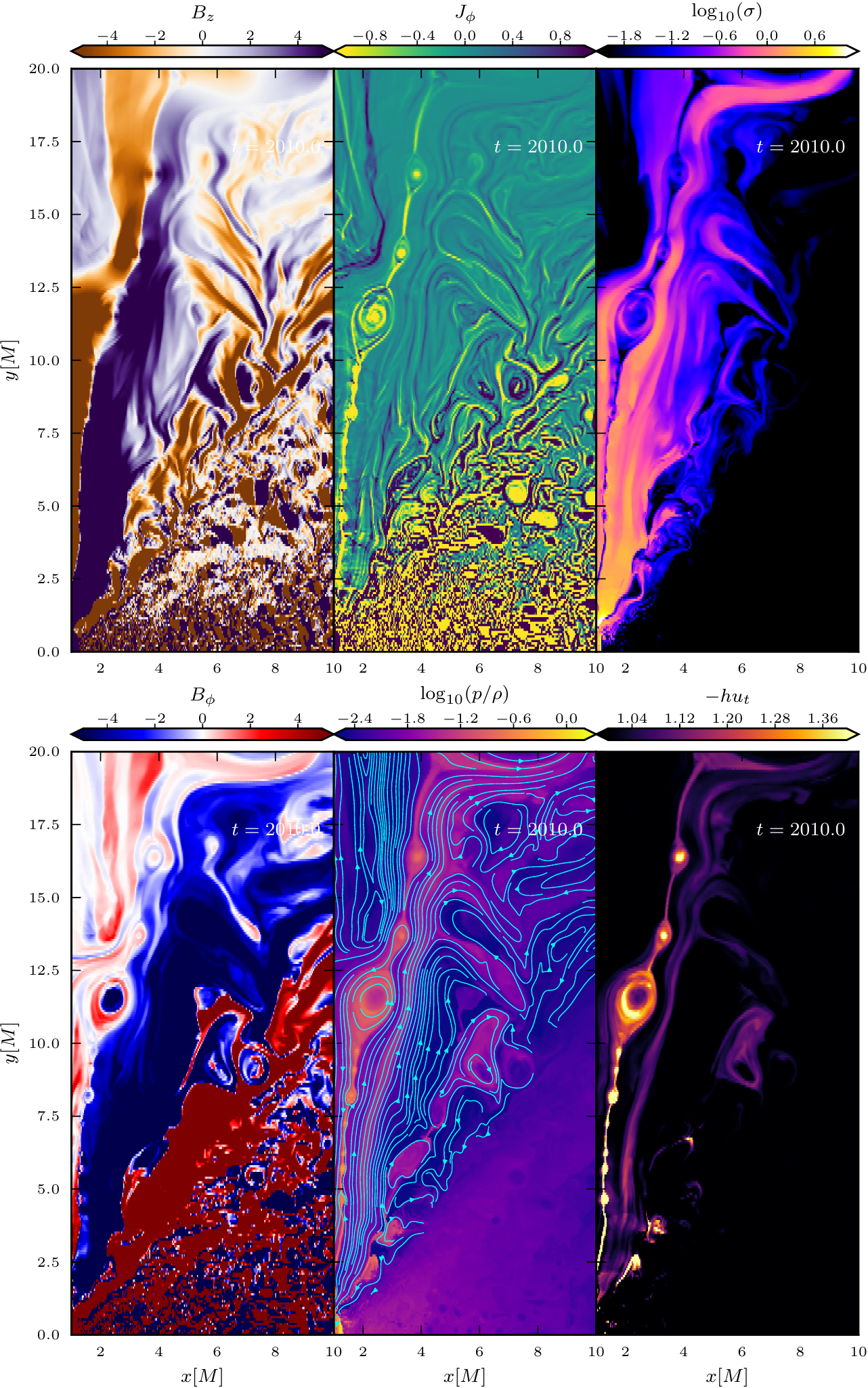}
  \end{center}
  \caption{Properties of the current-sheet structure formed in
    \texttt{\small{Model D}} at time $t = 2010 \, M$. First row, from
    left to right: $B_{z}$, $J_{\phi}$ and $log_{10}(\sigma)$, whereas
    in the second row: $B_{\phi}$, $p/\rho$, and $-h u_t$. In the middle
    panel of the second row, poloidal magnetic-field lines are
    illustrated in cyan (for clarity, magnetic-field lines are not shown
    in the dense torus region).}
    \label{fig:cs}
\end{figure*}

Because of the very large magnetization reached in models
\texttt{\small{Model A}} and \texttt{\small{B}} the fluid in these funnel
regions is expected to be accelerated and to become gravitationally
unbound. In order to ascertain whether this is the case, we employ the
\textit{Bernoulli criterion}, according to which a fluid element is
defined to be unbound if it has $hu_t\leq -1$, where $h$ is the specific
enthalpy of the fluid \citep{Rezzolla_book:2013}\footnote{A discussion
  between different definitions for the unbound criterion and their
  comparison has been discussed by \citet{Bovard2017}. Other contributions
  can be added to this criterion \eg magnetic and radiative
  \citep{Narayan2012,Chael2018}. The hydrodynamic prescription that we
  use effectively underestimates the amount of unbound/outgoing
  material, so that the material that we identify as unbound is actually
  going to reach a distant observer.}.

Figure \ref{fig:hut} reports the value of the quantity $hu_t$ for the
four models considered and clearly highlights that for
\texttt{\small{Models A}} and \texttt{\small{B}}, a large portion of the
material in the funnel is unbound and especially the one in the jet
sheath. On the other hand, when considering \texttt{\small{Models C}}
and \texttt{\small{D}}, it is clear that the unbound material is not
uniformly distributed, but appears in chains of plasmoids that move
outwards. In particular, while \texttt{\small{Model C}} exhibits a larger
number of out-going unbound plasmoids, which could be responsible of a
subsequent flaring activity, \texttt{\small{Model D}} shows fewer
plasmoids which are also accompanied by a uniform ejection of
matter. These distinct features will be further explored in the coming
sections.

\subsection{Magnetized regions of alternating polarity} 
\label{sec:lay}

One of the most useful quantities to monitor here is the toroidal
component of the magnetic field, which is initially zero but amplified
exponentially by the development of the MRI. In particular, and as we
will comment later on, the changes in polarity that the toroidal field
experience in regions of low rest-mass density -- or, equivalently, in
regions of high magnetization -- are closely connected to reconnection
and hence to the generation of plasmoids.

%% This can give the places that the magnetic field is changing polarity
%% and magnetic reconnection is expected. As we will see later, all
%% distinct features from \texttt{\small{Model C}} and \texttt{\small{D}}
%% come exactly from these surfaces where the sign of the toroidal
%% component is changing.

To highlight the role played by the toroidal magnetic field, we show In
Fig. \ref{fig:Bphi} the toroidal component of the magnetic field,
$B_{\phi}$. The rapid change of polarity, which is highlighted in the
figure by adjacent regions in blue/red, will lead to the formation of
current sheets. Figure \ref{fig:Bphi} shows quite remarkably how
\texttt{\small{Model A}} and \texttt{\small{Model B}} differ considerably
from \texttt{\small{Model C}} and \texttt{\small{Model D}}. The first
two, in fact, have large regions in the torus where the toroidal magnetic
field has the same polarity, while clear change of polarity lies at the
limit of the torus (this is similar to the current sheet discussed and
shown by \citealt{Ball2018}). This phenomenology is very different from
the one shown in the latter two models, where the toroidal magnetic field
inside the torus has a typical coherence lengthscale of a few $M$
only. More importantly, \texttt{\small{Models C}} and \texttt{\small{D}}
show that current sheets form also on the outer edges of the torus and
hence close to the jet sheath. This phenomenology is absent in models
\texttt{\small{A}} and \texttt{\small{B}}, and is responsible for the
generation and launching of the plasmoids\footnote{Also the interior of
  the torus is highly turbulent region with changes of polarities that
  have even smaller lengthscales. However, because of the high density in
  the torus and and the very low magnetization, the acceleration of
  plasmoids via reconnection is expected to be suppressed in these
  regions \citep{Li2017PIC}.}

%% Vanishing $B_{\phi}$ does not imply always the existence of a current
%% sheet. However in accretion flows, due to differential rotation, the
%% change of sign in the toroidal field implies a change of polarity in
%% the poloidal reconnecting field. As such a change of sign in
%% $B_{\phi}$ and a region that goes to zero does imply the occurrence of
%% a current sheet.

The presence of current sheets in the toroidal direction can also be
appreciated through the distribution of the toroidal current
$J_{\phi}:=(\nabla \times B)_{\phi}$, which we report in
Fig. \ref{fig:jphi} for the four models and at a representative time.
Note that for \texttt{\small{Model A}} and \texttt{\small{B}}, the
funnel regions are essentially devoid of current sheets and exhibit
almost constant and essentially zero toroidal currents. This is in
contrast to what happens for \texttt{\small{Model C}} and
\texttt{\small{D}}, whose funnel regions show highly variable regions of
toroidal currents with changing polarities.

Combining the information from Figs. \ref{fig:Bphi} and \ref{fig:jphi}
with that on the unbound material in Fig. \ref{fig:hut}), it appears
clear that as consecutive loops of alternating polarity are advected from
the torus near the black hole, current sheets are constantly generated
above and below the black hole. The conversion of magnetic energy into
internal energy result from the reconnection, accelerates this material
up to energies that make it become unbound.

%% In the bulk of the torus where the density is very high and the field is
%% changing also significantly in very small volumes we do not show the same
%% contours. In this highly turbulent region, due to the high density and
%% the very low magnetization any acceleration is expected to be suppressed
%% \citep{Li2017PIC}. For \texttt{\small{Model A}} and \texttt{\small{B}} a
%% clear change of polarity lies at the limit of the torus, this is similar
%% to the current sheet discussed and shown in \citep{Ball2018}.

%% For the next two models, \texttt{\small{Model C}} and \texttt{\small{D}},
%% the whole funnel region above the black hole is filled with current
%% sheets. All these regions of alternating polarity give rise to magnetic
%% reconnection and to the unbound features shown in fig. \ref{fig:hut}. The
%% consecutive loops of alternating polarity from the torus are constantly
%% advected through accretion to the vicinity of the black hole, this
%% process creates constantly several current sheets above and below the
%% black hole. Material heats up in these regions and becomes unbound
%% (fig. \ref{fig:hut}).

In order to have a closer look on such a region, we focus in
Fig. \ref{fig:cs} on \texttt{\small{Model D}} at time $2000 M$, where
current sheets are clearly visible and easily identified. More
specifically, the six panels of the figure report from left to right:
$B_{z}$, $J_{\phi}$, $log_{10}(\sigma)$, $B_{\phi}$, $p/\rho$, and
$-h u_t$. Poloidal magnetic-field lines in the middle panel of the
second row panels, although such lines are not shown in the dense torus
region for clarity.
%% Regions of alternating magnetic-field polarity are closing up,
%% brought to the low density region, dragged from their footpoints due to
%% accretion from to the torus. In the third panel of Fig. \ref{fig:cs} we
%% show $B_{\phi}$ the toroidal component of the magnetic field.
Overall, the various panels show that when going from the equatorial
plane towards the polar axis, regions of different polarity appear both
in the poloidal and in the toroidal magnetic field. Due to the accretion
process, these regions are progressively pushed together and forced to
reconnect in different sites. The reconnection layers that lie inside the
torus can contribute to electron heating, but because the magnetization
in such regions is very low, this material is not expected to contribute
to any flaring activity. On the other hand, the reconnection layers in
the polar (funnel) region, where the density is low and the magnetization
is high, can be expected to be accelerate and unbind matter more
efficiently (see last panel of the second row in Fig. \ref{fig:cs}) and
hence represent the optimal sites to produce a flaring activity.

In the central panel of the bottom row, we report the ratio $p/\rho$, as
a guide to distinguish regions of different temperature and over-plotted
in cyan, are the poloidal magnetic-field lines. Combining information
from the poloidal and toroidal magnetic-field strengths it is possible to
reconstruct the various regions where magnetic-field lines change
polarity, thus indicating the presence of a reconnection layer. These
layers can also be tracked through the azimuthal current $J_{\phi}$
(central panel of the upper row in Fig. \ref{fig:cs}). Plasmoids are
generically formed in these reconnection layers, but a clear distinction
should be made between those that eventually become unbound and the ones
that do not leave the black-hole--torus system. In particular, in regions
with high magnetization $\sigma$ and high Alfv\`en speed (right panel in
the upper row of Fig. \ref{fig:cs}), the process of acceleration is more
efficient and plasmoids generated in such regions are energized becoming
unbound (right panel in the lower row of Fig. \ref{fig:cs}). On the other
hand, if the magnetization is not sufficiently high, the plasmoid
produced fail to be launched, as shown in the in the upper-left corners
of the various panels in Fig. \ref{fig:cs}, where a reconnection layer
can be clearly seen in the panels for $J_{\phi}$ and $B_{\phi}$, but not
in the panel for the Bernoulli constant $-h u_t$. Since the magnetization
in this region is almost two orders of magnitude smaller, the efficiency
of the conversion of magnetic energy to internal energy is much smaller
and hence does not lead to a plasmoid launching. This is also reflected
by the absence in this region of an extra heating and can be clearly
appreciated from the central panel in the lower row of Fig. \ref{fig:cs},
whose upper-left corner is about one order of magnitude colder.
 
Figure \ref{fig:cs} shows that there are several plasmoids of various
sizes that can be clearly found across the six panels. However, the large
magnetic field in the main reconnection layer in the central region of
Fig. \ref{fig:cs} (\ie at $1\,M\lesssim x \lesssim 4 \,M$, $2.5\, M
\lesssim y \lesssim 20 \,M$), one able to produce and energize a whole
series of plasmoids along the current sheet, \ie a ``plasmoid chain''. By
looking at the magnetization parameter (top right panel) we see that main
reconnection layer responsible for the plasmoid chain lies in a region
with high magnetization where $\sigma \approx 3$ and is never smaller
than $\sigma \geq 0.3$. This is in agreement with the results of
\citealt{Li2017PIC, Ball2018a}, and highlights that reconnection layers
with low magnetization, i.e., $\sigma \leq 0.0 1$, are not efficient in
producing highly energized pockets of plasma.

It is also worth remarking about the reconnection layers that are
developed in \texttt{\small{Model A}} between the torus and the
low-density regions above and below the torus. These are clearly visible
in Fig. \ref{fig:Bphi} and \ref{fig:jphi} (leftmost panels), where
reconnection layers of alternating polarity develop further out from the
funnel-jet region. The reconnection processes taking place in these
regions could contribute to the heating of the local material. Also in
this case, however, we do not expect these plasmoids to contribute to any
flaring activity because of the low magnetization in these regions, which
is $\sigma \lesssim 0.01$.

Before closing this section, we briefly discuss the conditions for the
activation of the \cite{Blandford1977} (BZ) mechanism to power the jet
and extract energy from the rotating black hole. Following
\citet{Komissarov:2009dn}, the critical condition for the activation of
the BZ mechanism is for the Alfv\`en velocity to exceed the free-fall
velocity at the ergosphere. This condition, which is not straightforward
to validate, effectively translates into requiring that the magnetization
is $\sigma >1$ at the ergosphere. Furthermore, this condition is clearly
always fullfilled for \texttt{\small{Models A}} and \texttt{\small{B}},
for which the magnetization in the jet is always very high, but this is
less clear for \texttt{\small{Models C}} and \texttt{\small{D}}, for
which $\sigma \lesssim 1$, close to the black hole, thus implying that
the BZ mechanism could be quenched at times. Conversely, there are times
in the dynamics of models \texttt{\small{Models C}} and
\texttt{\small{D}} when the funnel region accumulates enough magnetic
flux of a single polarity to exceed $\sigma > 1$.
%
%% As such the magnetization $\sigma$ of the funnel region indicates
%% whether the BZ mechanism is activated and as a consequence a
%% production of a jet or not.
%% The \textit{funnel wall}, is usually demarcated by $\sigma=1$, indicating
%% that regions with lower magnetization do not contribute to the jet
%% production. Close to the \textit{funnel wall} (usually encloses it) is
%% the boundary of bound material to the disk, which means that the whole
%% magnetized funnel is unbound.
This is shown, for example, in Fig. \ref{fig:sigma}, which shows that the
upper funnel of \texttt{\small{Model C}} and the lower funnel of
\texttt{\small{Model D}} are both highly magnetized, thus with a
potentially active BZ mechanism. However, because the accumulated
magnetic flux varies continuously and magnetic field with opposite
polarities is regularly accreted, it is difficult to build a steady jet
and hence generate the physical conditions necessary for the development
of a stationary BZ mechanism. This behaviour is also reflected in
Fig. \ref{fig:mdot} and in particular in the upper panel, which clearly
shows how the jet efficiency if always less than unity in
\texttt{\small{Models C}} and \texttt{\small{D}}.

%-----------------------------------------------------------------------
\section{Dynamics of the plasmoids} 
\label{sec:plasmoids}
%-----------------------------------------------------------------------

Thus far we have described models where reconnection layers naturally
develop as a result of the initial magnetic-field configuration embedded
in the torus. These reconnection layers result in the formation of
plasmoids and plasmoid chains. These quasi-spherical blobs are sometimes
energized and eventually outgoing. However, this is not always the
case. Other times some of these blobs are either advected and accreted by
the black hole or are bound to the torus. Since unbound plasmoids are
conjectured to be filled with high-energy particles and may be at the
origin of flares in AGNs \citep{Yuan2009, Giannios2013b, Younsi2015, 
Li2017flare}, it
is interesting to track their evolution both in terms of kinematics, (via
spacetime diagrams) but also in terms of their of their thermodynamics
(via the evolution of the magnetization and temperature). Note that this
interest goes beyond a mere question of plasma dynamics, since, as
discussed by \citet{Younsi2015}, the evolution of the physical properties
of a plasmoid is essential to assess its role and contribution to a
flare. For compactness, we postpone a more detailed discussion of this
point to a forthcoming future study.

Finding and tracking plasmoids is however far from trivial, especially if
one wants to rely on a fully automated identification and classification
pipeline. To this scope, we next describe two different ways to identify
individual plasmoids and track them during their evolution. A more
detailed discussion on the tracking procedure can be found in the
Appendix \ref{appen:track}. To this scope, hereafter our attention will
be focused on a specific large plasmoid that develops in
\texttt{\small{Model D}} at time $t\approx 5000\, M$.

\begin{figure*}
  \begin{center}
    \includegraphics[width=0.31\textwidth]{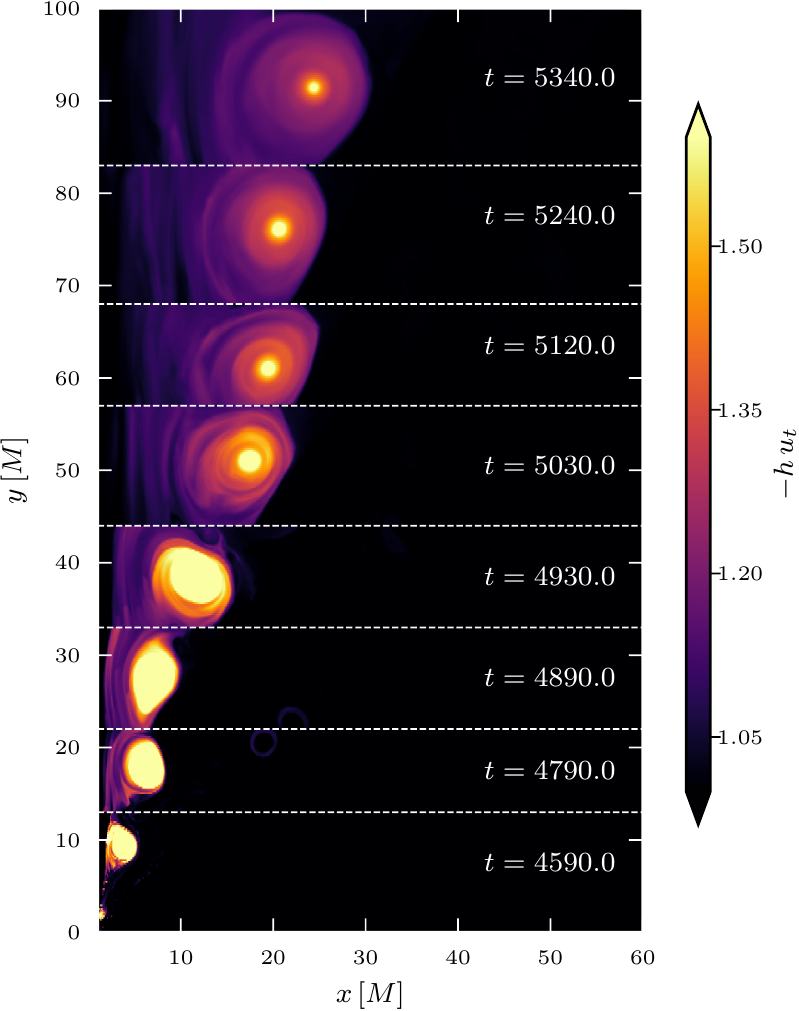}
    \hskip 0.2cm
    \includegraphics[width=0.31\textwidth]{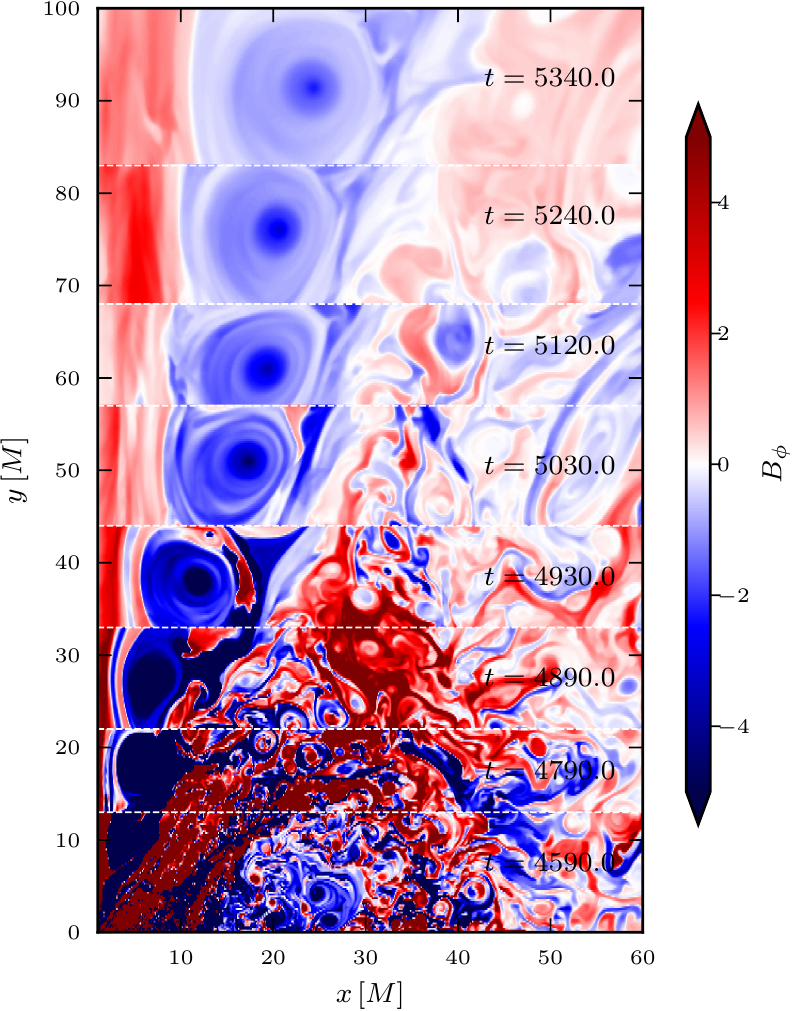}
    \hskip 0.2cm
    \includegraphics[width=0.32\textwidth]{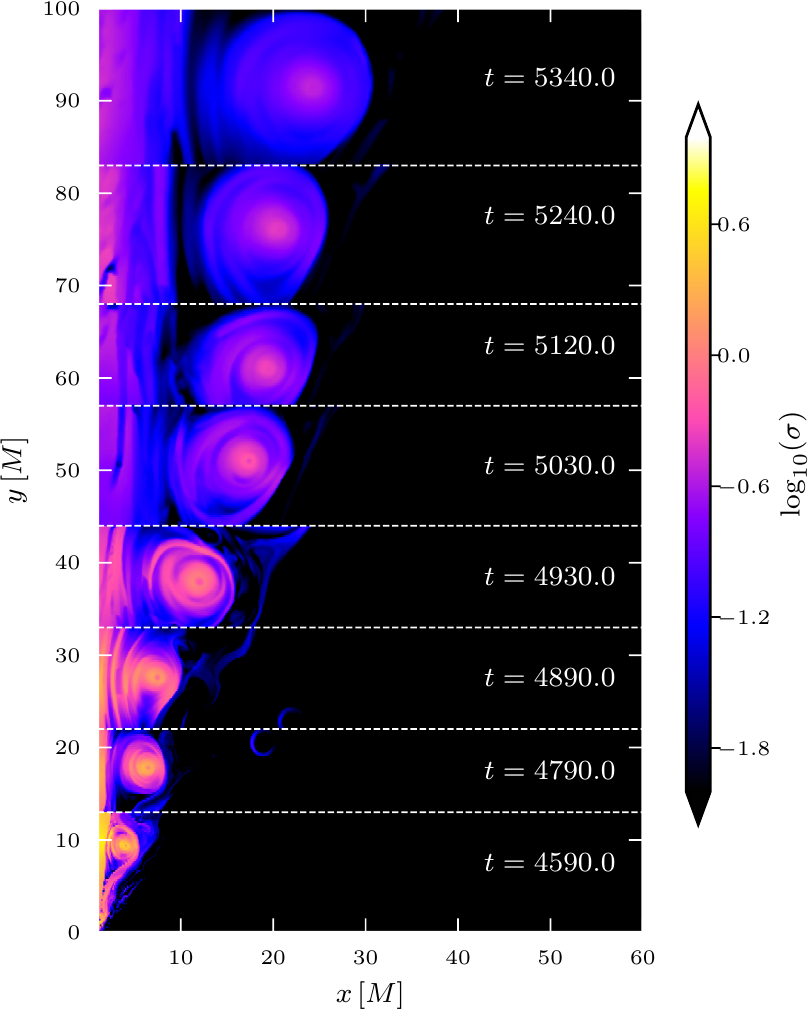}
  \end{center}
  \caption{Spacetime diagram showing the evolution of a detected plasmoid
    from \texttt{\small{Model D}}, through the Bernoulli constant $-h
    u_t$, the $B_{\phi}$ and the magnetization parameter $\sigma$ 
    at times from $t = 4590\, -\, 5340 \, M$.}
    \label{fig:hut_8}
\end{figure*}

\begin{figure}
  \begin{center}
    \includegraphics[width=0.75\columnwidth]{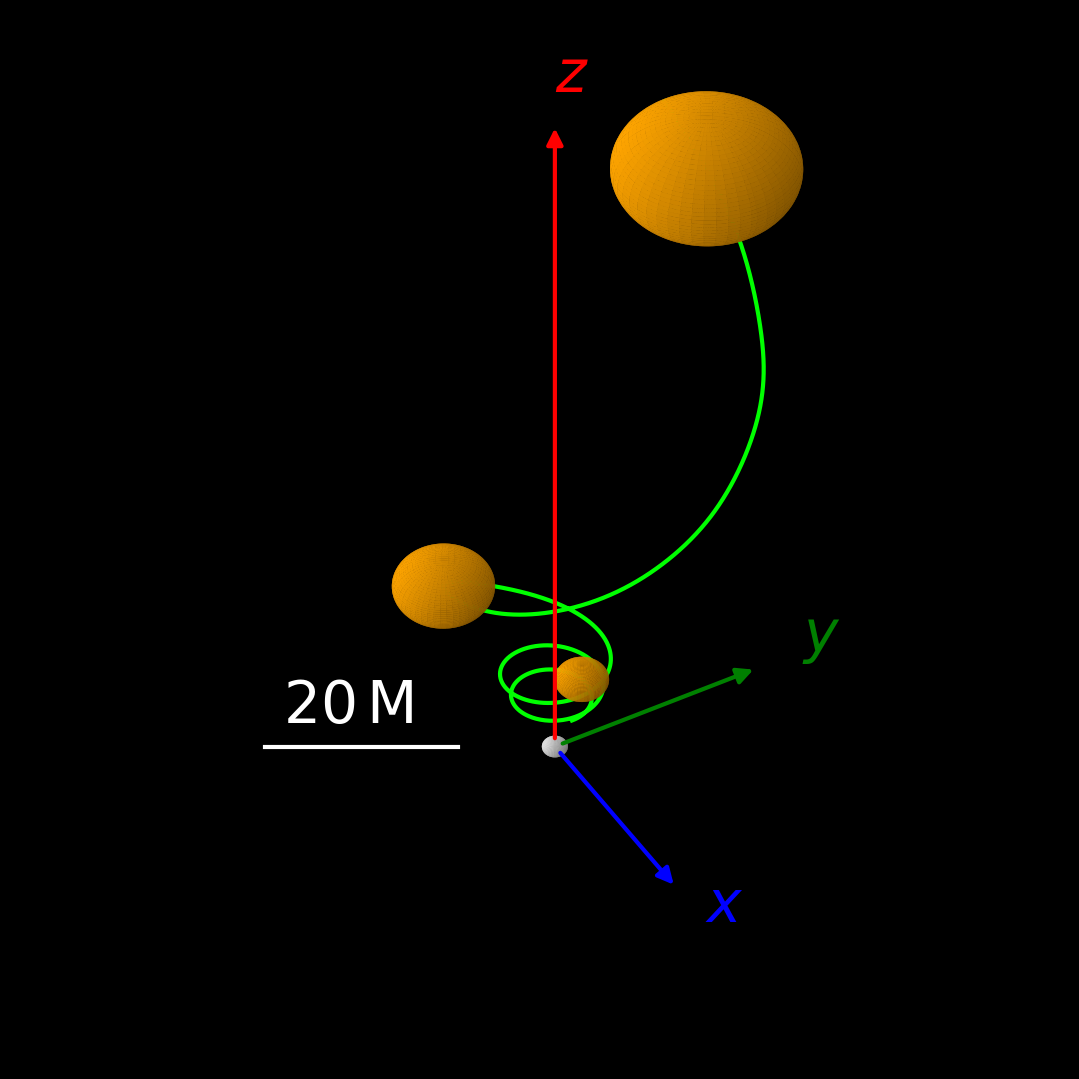}
  \end{center}
  \caption{3D reconstruction of the trajectory of the outward moving
    plasmoid shown in Fig. \ref{fig:hut_8} after integrating in time the
    velocity of its core, including the azimuthal component.  }
    \label{fig:track}
\end{figure}

\begin{figure*}
  \begin{center}
    \includegraphics[width=0.65\textwidth]{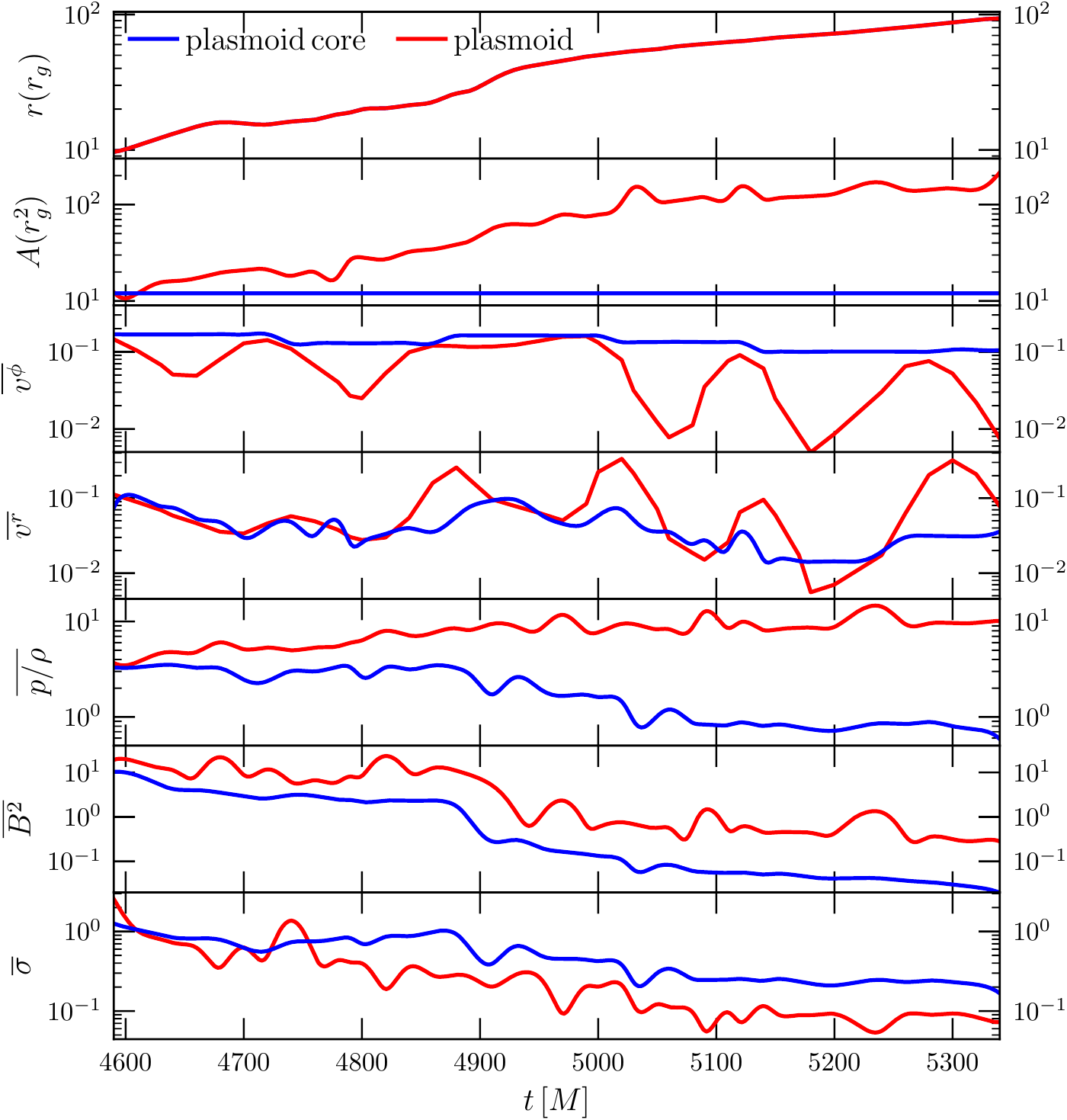}
  \end{center}
  \caption{Evolution of a detected plasmoid from \texttt{\small{Model
        D}}, from $t = 4590\, -\, 5340 \, M$. The evolution of the whole
    plasmoid is shown in red and the evolution of the core of the
    plasmoid in blue. The first panel shows the distance of the plasmoid 
    from the black hole, whereas the second panel the evolution of the area
    that encloses the plasmoid, which for the core of the plasmoid is
    constant. The next five panels show the average $\sigma$, $p/\rho$,
    $B^2$, $v^{\phi}$ and $v^r$ respectively, for the plasmoid and its
    core. The last panel shows the distance of the core of the plasmoid
    from the black hole.}
    \label{fig:plasm_6}
\end{figure*}

\begin{figure*}
  \begin{center}
    \includegraphics[width=0.75\columnwidth]{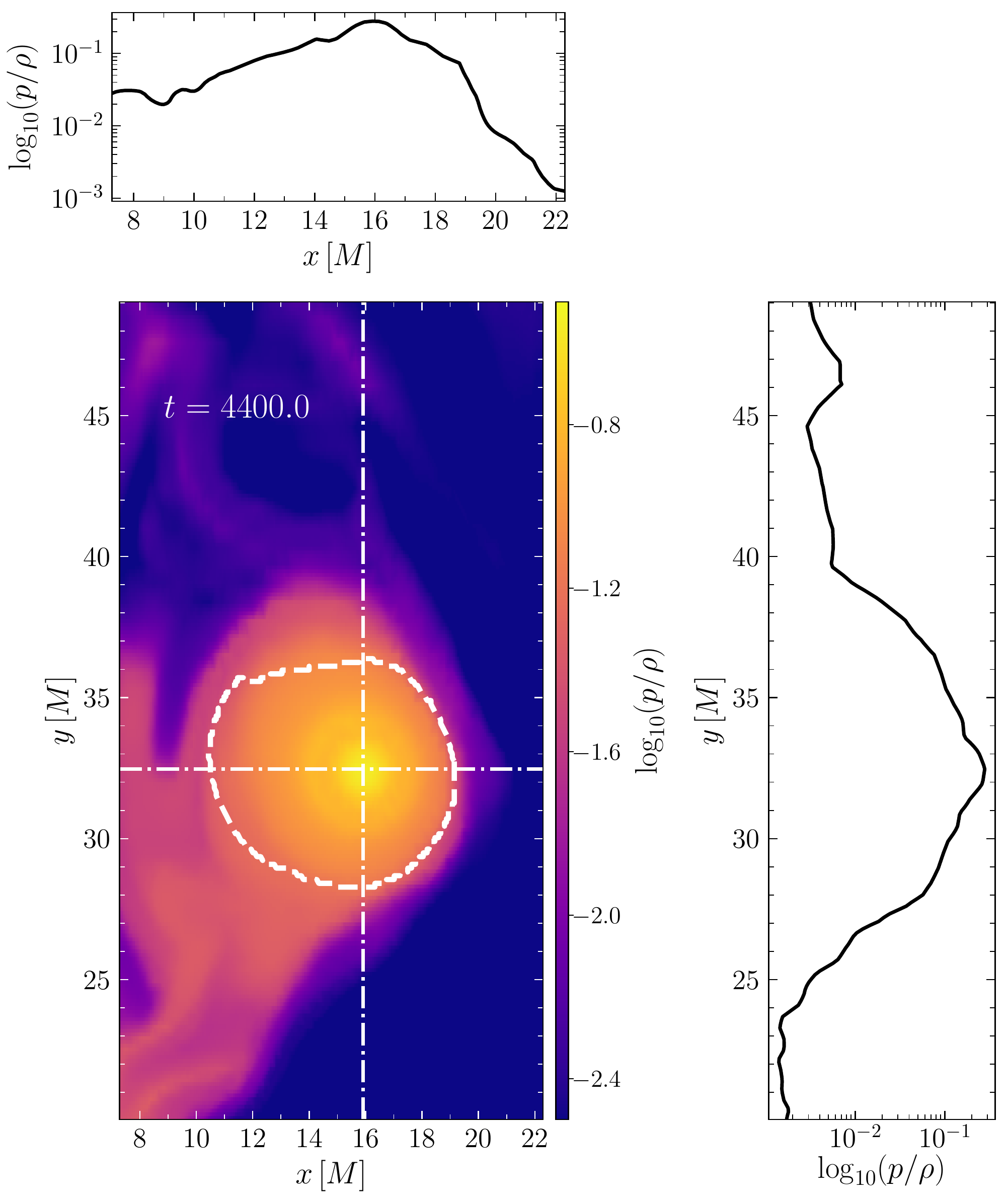}
    \hskip 1.5cm
    \includegraphics[width=0.75\columnwidth]{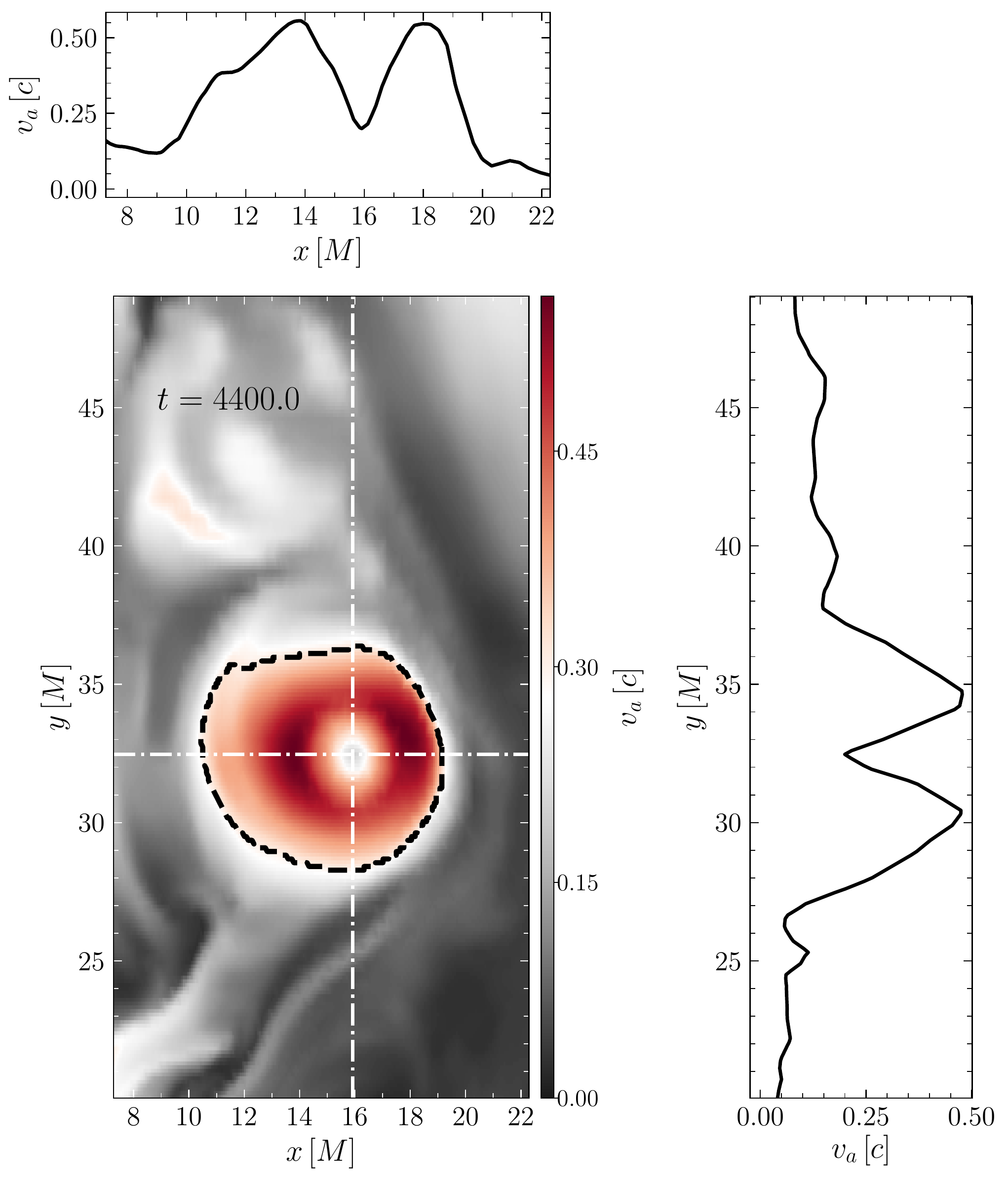}
  \end{center}
  \caption{Detection of a plasmoid from \texttt{\small{Model D}} for
    either the temperature proxy (left panels) or the Alfv\`en speed
    (right panels). The two (horizontal and vertical) dashed lines pass
    though the center of the plasmoid. The two subplots show exactly the
    cuts made by these lines through the plasmoid.}
    \label{fig:cutpr}
\end{figure*}

The two methods are based on the different properties of the outgoing
plasmoids, with the first method being based on the Bernoulli constant,
whereas the second one on magnetic properties such as $\sigma$, the
Alfv\`en speed $v_a$, and the magnetic energy. Due to the essentially
circular shape in which plasmoid are produced (see Figs.
\ref{fig:sigma}, \ref{fig:hut} and \ref{fig:cs}), we search the data for
plasma regions that have such a shape and that, additionally, are
characterised by a non-negligible magnetization, \ie $\sigma \geq 0.3$
and with a large Bernoulli constant, \ie $-h u_t \ge 1.2$. The use of
these cut-offs in the magnetization and binding energy essentially remove
from our search the whole torus, but also considerable portions of the
regions above and below it.

Leaving the details to Appendix \ref{appen:track}, below are briefly
listed the main steps we have followed in detecting and tracking plasmoids:
\begin{enumerate}
\item detect a plasmoid using the Bernoulli constant $-h u_t$ and then
  the difference of two images produced deconvolving with a Gaussian
  kernel of increasing standard deviation to identify bright blobs (this
  is also known as the ``difference of Gaussians'' method).
\item isolate a squared region of plasma including the center of the
  plasmoid found and with size set by the radius estimate from the
  previous step.
\item apply the Canny-edge detector algorithm to the Bernoulli constant
  distribution so as to determine the boundary of the plasmoid.
\item if the result of the previous step is not one closed curve, we use
  the resulting curve/curves from the previous step to find its convex
  hull (\ie the smallest convex set comprising the plasmoid).
\end{enumerate}
Most of the time, the detection procedure ends successfully at step (iii)
and in this way we have been able to isolate and track accurately the
plasmoid in Fig. \ref{fig:hut_8}. While this procedure could be applied
to determine also other plasmoids and thus explore their statistical
properties, it is also computationally intensive and we therefore decided
to postpone this investigation to a subsequent work. At the same time,
since the detected plasmoids are produced in regions with high
magnetization, we assess the impact of the numerical resolution on the
statistics of plasmoid production by measuring the volume fraction of
regions of a given magnetization (\ie with $0.01\leq \sigma \leq 10$)
over a time window of $1000\, M$ and study how this changes with
resolution. In this way, we find the volume fraction of highly magnetized
regions (\ie with $1.2\lesssim \sigma \lesssim 10.0$) does not change
considerably with resolution (see discussion in Appendix \ref{appen:res}
and Fig. \ref{fig:hist_sigma} for a visual impression).

Particularly interesting are of course the largest plasmoids as these are
the most energetics and hence directly related to a possible flaring
activity in AGNs; besides, the largest plasmoids are also those that are
not significantly affected by the chosen resolution (see Appendix
\ref{appen:res} for a discussion). Once a plasmoid is found, it can be
tracked in its time evolution as shown in Fig. \ref{fig:hut_8}, which
reports in a spacetime fashion the kinematics of a large and unbound
plasmoid in its outward motion. More specifically, the left and right
panels of Fig. \ref{fig:hut_8} report the appearance of the plasmoid in
terms of the Bernoulli constant (which is particularly effective in
detecting the plasmoid) and of the magnetization (which is particularly
effective in determining the boundary of the plasmoid), respectively [In
  all snapshots except one, the boundary is found at step (iii)]. In the
middle panel of Fig. \ref{fig:hut_8}, on the other hand, we show the
spacetime evolution of the plasmoid in terms of the toroidal magnetic
field, highlighting its motion from regions of very strong toroidal
magnetic field near the black hole, over to much weaker areas near the
jet.

Note that as the plasmoid leaves the central regions near the black hole
and moves outwards in regions of decreasing density and pressure, it
expands because of the increased internal energy and accelerates moving
outwards. Assuming the plasmoid is quasi-spherical and hence has a
limited extent in the azimuthal direction, the reconstructed plasmoid
trajectory is obtained after integrating in time the velocity of its
core, including the azimuthal component. This is shown in
Fig. \ref{fig:track} where it is seen that the expanding plasmoid
performs $\sim 3$ orbits from its formation close to the black hole at
$(4,10)\, M$ to the end of the track at $(25,100)\, M$. When scaled to
the mass of Sgr~A*, the orbital period of the helical trajectory ranges
from $30\,\rm min$ to $7\,\rm h$. It is an intriguing possibility to
consider the orbiting plasmoid discussed here in the context of the
astrometrically resolved flares recently observed by the GRAVITY
collaboration \citep{Abuter2018}.

A more quantitative assessment of the dynamical and thermodynamical
evolution of the plasmoid is presented in Fig. \ref{fig:plasm_6}, where
we show several quantities relative to the plasmoid, either when
spatially averaged or when referring to the core of the plasmoid, which
we define as a circle of radius $2\, M$ centered at the plasmoid's
center.  In the upper panel of Fig. \ref{fig:plasm_6}, we monitor the
distance of the plasmoid from the black hole, whereas in the second row
we show the evolution of the size of the plasmoid by measuring its
surface area. Note that the plasmoid's size increases continuously,
reaching a size of $215\, r_g$ after $\approx 800\, M$ from its
formation. Through the third and fourth row of Fig. \ref{fig:plasm_6},
where we plot the azimuthal component and the radial velocity, we can
reconstruct the plasmoid's kinematics.  Note that the average velocity of
the plasmoid fluctuates considerably (red line in Fig. \ref{fig:plasm_6})
and is quite distinct from the evolution of the average velocity of the
core of plasmoid (blue line). More specifically, while the latter remains
almost constant at a velocity of $\simeq 0.1$, the former varies
quasi-periodically, oscillating between $\simeq 0.01$ and $0.2$.

Another important quantity characterising the properties of the
plasmoid
is its temperature, which we measure through the ratio
$p/\rho$. Interestingly, the core of the plasmoid cools significantly,
namely, by almost an order of magnitude over the time the plasmoid is
followed, essentially because of the decrease in pressure as it moves
outwards. On the other hand, the whole plasmoid's temperature increases
steadily over time as a result of the interaction of the plasmoid with
the surrounding matter and its dynamics. Finally, the average
magnetic-field strength (sixth row) is decreasing, together with the
magnetization (seventh row). 

Any single plasmoid that can be isolated and tracked, can then be studied
in detail in all of its interesting quantities. As an example, we report
in Fig. \ref{fig:cutpr} one-dimensional cuts for a single plasmoid and in
terms of the quantities $p/\rho$ and $v_a$. The two subplots show the
vertical and horizontal cuts through the center of the plasmoid. In the
right panel of Fig. \ref{fig:cutpr}, the shape of the plasmoid (which
remains unchanged when shown in terms of the magnetization) provides a
representative example about the use of the second method for the
detection of a plasmoid, in which we first determine the center of the
plasmoid as a local minimum and then detect the boundary of the plasmoid
by finding first the local maxima. At the same time, the same panel shows
how $p/\rho$ falls off by two orders of magnitude when reaching the outer
layers of the plasmoid. A similar structure is evident also in the
Bernoulli constant $-h u_t$, as is seen in the right panel
Fig. \ref{fig:cutpr}.

%-----------------------------------------------------------------------
\section{Conclusions} 
\label{sec:con}
%-----------------------------------------------------------------------

We have performed a series of GRMHD simulations of accreting tori onto a
rotating black hole employing different topologies for the initial
magnetic field.
One goal of these simulations was to
illustrate and confirm that when employing an initial magnetic-field
configuration that does not consist of the standard single poloidal-field
loop, a stationary jet configuration does not form.

%% \op{There seem to be dramatic differences between model B and model
%%   C although only the polarity is flipped.  Model B behaves like a
%%   one-loop, model C has much more small scale features and plasmoids.
%%   This should be discussed in the conclusions too.  How about this
%%   paragraph here:}

We have observed that the evolution of toris with initial magnetic loops
with alternating polarity (multi-loop initial topology;
\texttt{\small{Models C}} and \texttt{\small{D}}) differs dramatically
from the dynamics of small scale loops of the same polarity
(\texttt{\small{Model B}}). Same-polarity loops quickly reconnect to form
a large loop with resulting flow similar to the single loop setup
(\texttt{\small{Model A}}). On the other hand, opposite-polarity loops
preserve their small coherence length, giving rise to copious plasmoids
in the ensuing turbulent evolution.  These magnetic fields with small
coherence lengths are advected to the black hole and give rise to
fluctuating jets of low average power ($10^{-4}-10^{-3}$ times the jet
power with tori having the same polarity) but large variability.  On rare
occasions, the fluctuating jet power can reach or exceed the typical jet
power in the same polarity cases which amounts to $1-10\%$ of the
accretion power in our \texttt{\small{Model A}} scenario. In addition,
tori having a multi-loop initial topology produce an accretion flow that
does not produce a stationary magnetized jet, but a series of regions of
poloidal and toroidal magnetic fields with alternating polarities. At the
boundaries of these regions, reconnection can take place and generate
plasmoids with large magnetization.

%\lr{have you computed the jet power
%  in model A and model D? How much do they differ}
%  \an{The jet for all models (including A and D) are shown in the 
%  middle panel fo fig. \ref{fig:eta_plasm}. They differ 3-4 orders 
%  of magnitude.}
%
%% \an{I agree, however lets call replace SANE with Model A, since we 
%% have not introduced the terminoly SANE anywhere.}
%
%% \an{The absence of a stable jet, when employing B-field configurations
%% other than nested loops, have been discussed in other studies. I would
%% not call this our ultimate goal.}
%
Most of the plasmoids are either accreted to the black hole or remain in
the high-density torus. However, those plasmoids generated in reconnection
layers with relatively high magnetization, \ie $\sigma \gtrsim 0.3$, are
outward moving in the funnel region and gravitationally unbound. 

While the appearance of plasmoids generated during the simulations is
rather straighforward, their automatic detection and characterization
is far more complex. To handle this processs we have devised two
different methods to detect the plasmoids at any given time during the
simulation. In essence, we detect the center of a plasmoid in terms of
a ``blob-detector'' algorithm and then define the boundary of the
plasmoid either as an edge (usually in terms of the Bernoulli constant
$-h u_t$ or of the temperature $p/\rho$) or as local maxima (usually
in terms of the Alfv\`en velocity $v_a$ or the magnetization
$\sigma$). In this manner, plasmoids can be tracked in the most
interesting regions of the domain and their dynamics studied
individually.

Recent observations of flaring activity and variability from AGNs hint
that very rapid particle acceleration is required and that the
emission region is relatively compact, few Schwarzschild radii
\citep{Levinson2007, Begelman2008, Ghisellini2008, Giannios2009}.
Magnetic reconnection in the vicinity of a black hole can provide both
rapid particle acceleration and a compact emission region. Our
simulations have therefore shown that the models with an initial
(opposite polarity) multi-loop magnetic geometry exhibit an intense
variability in the power lost via outflows. This is ultimately the
result of the accretion process, which forces small scale loops to be
transported near the black hole -- in particular in the polar regions,
where the magnetization is larger -- and reconnect. The consequent
release of magnetic energy in these plasma regions and the formation
of plasmoid are then responsible for the intense variability in the
emitted power. Hence, the generic behaviour found in these simulations
can have implications on the observed variability of AGNs, even when
they are experiencing low accretion rates, such as Sgr~A$^*$.

Furthermore, the episodic reconnection that occurs close to the black
hole produces plasmoid chains filled with relativistic particles. These
plasmoids can represent an additional source of variability in the
vicinity of the black hole. The production of large plasmoid chains also
has a considerable impact on the emitted power of the outflow, which is
observed to increases of more than two orders of magnitude at times
\citep{Ponti2017, Do2019}. Finally, the evidence that these plasmoids
have nonzero angular momentum and hence an orbital motion, makes them
potentially related to the observations of orbiting material recently
made near the galactic center \citep{Abuter2018}.

Despite the absence of physical resistivity in our mathematical
formulation of the GRMHD equations, magnetic reconnection is produced
during the simulations and is generated entirely by the finite numerical
resolution. As a result, the numerical methods employed in this study
could be considered as inadequate for a detailed description of the
generation and evolution of energetic plasmoids. However, as shown
through the comparison of simulations carried out at different and
increasing resolutions (see Appendix \ref{appen:res}), the main
qualitative features discussed here are robustly produced across all
different resolutions. This provides convincing evidence that the basic
resistive features discussed here are qualitatively correct and that
reconnection occurring in magnetically dominated regions can produce
energized plasmoids that are outgoing and gravitationally unbound.

There are several and natural extensions of this work. First, by carrying
out a more detailed investigation of the statistical properties of the
plasmoids. This involves not only the processes leading to their
formation, but also to the factors that determine their evolution and
emission. Second, and more importantly, by investigating the
phenomenology discussed here when employing a fully resistive formulation
of the GRMHD equations (see, \eg \citealt{Palenzuela:2008sf,
  Dionysopoulou:2012pp, Ripperda2019}), so as to assess the properties of
the plasmoids when different and resolution-independent values of the
physical resistivity are considered. A complete description of the MHD
properties of the plasmoids will allow the study of particle acceleration
and radiation signatures due to the plasmoid evolution in such a
turbulent environment around black holes \citep{Bacchini2019}.

%-----------------------------------------------------------------------
\section*{Acknowledgements}
%-----------------------------------------------------------------------
  
AN is supported in part via an Alexander von Humboldt
Fellowship. Additional support comes in part from HGS-HIRe for FAIR, the
LOEWE-Program in HIC for FAIR, ``PHAROS'', COST Action CA16214, the ERC
Synergy Grant ``BlackHoleCam: Imaging the Event Horizon of Black Holes''
(Grant No. 610058). The simulations were performed on SuperMUC at LRZ in
Garching, on the GOETHE-HLR cluster at CSC in Frankfurt, and on the
HazelHen cluster at HLRS in Stuttgart.

\begin{figure*}
  \begin{center}
    \includegraphics[width=0.75\textwidth]{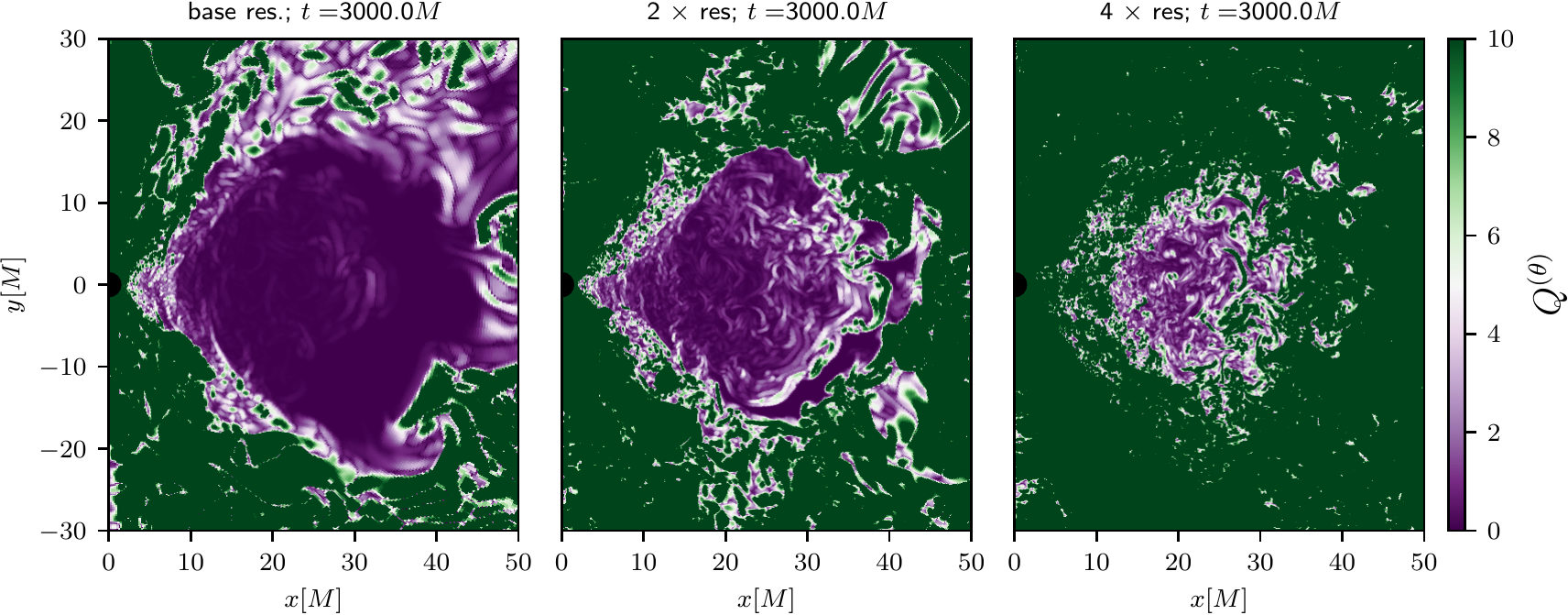}
  \end{center}
  \caption{The MRI quality factor $Q_{\theta}$ for \texttt{\small{Model
        D}} for the three different resolutions $1\times$, $2\times$, and
    $4\times$ the base resolution.}
    \label{fig:Q}
\end{figure*}

\section*{}
\bibliographystyle{mnras}%{aasjournal}    %{mnras}
\bibliography{aeireferences}

\appendix
%-----------------------------------------------------------------------
\section{Plasmoid Tracking}
\label{appen:track}
%-----------------------------------------------------------------------

In this Appendix we discuss in more detail the methods we use to detect
and track plasmoids and follow their evolution. The study of their
dynamics, expansion and cooling, represents a first step towards modeling
flares emerging from the vicinity of the black hole in AGNs.

More specifically, in tracking the plasmoid we have made use of the
python package \texttt{scikit-image} \citep{scikit}. In order to identify
blob structures, we use the difference of Gaussians method
\citep{Herbert2006}. In this approach, the data-image is successively
convolved with a Gaussian kernel with increasing standard deviation
(essentially blurred). The difference between the two new images is used
to identify higher-intensity regions against a lower-intensity
background. In this first step, the identification is done through the
Bernoulli constant $-h u_t$, since it has turned out to yield more robust
results. After a plasmoid is identified, the coordinates of its center
are determined. Usually, this comes with an estimation of the radius of
the plasmoid. However, most of the plasmoids during the simulation are
not completely spherical, so the radius alone cannot define the boundary
of the plasmoid. In most cases, the plasmoid-structure has several
irregularities, so that we divide the domain into squares whose number is
proportional to the number of plasmoids that have been found. For every
plasmoid, we cut a neighbourhood around it that includes the radius
estimation from the previous step. In this way, in every square there is
only one plasmoid, whose boundary is found using the Canny edge-detector
algorithm \citep{Canny1986}. The algorithm is applied
%% \lr{I would remove
%%   the text in green as it is not particularly intereesting or clear}
%% \an{ok} \textcolor{green}{In the first method we continue to search for
%%   the edges through the same parameter $-h u_t$, $p/\rho$ yields similar
%%   (if not identical) results. Again in this step, every image is
%%   convolved with a Gaussian to reduce the effect of noise. Then by
%%   evaluating the gradient of the intensity of the image, potential edges
%%   are found. Subsequently, by removing non-maximum points a curve is
%%   found, which is the boundary of the plasmoid.}  
In the case when the plasmoid has an irregular boundary, we are left with
a curve which is not closed or with two or more smaller curves. In this
case, then we continue by finding the convex hull of the resulting curve
or curves from the previous step. In all tests, at this point the
boundary of the plasmoid was adequately described.

In the second method used, after detecting the plasmoid with the same way
as discussed before, we then cut the domain into squares that include the
center of the plasmoid and its radius estimation. Next, we focus on the
magnetization parameter (or the Alfv\`en speed), which normally exhibits
a local minimum at the center of the plasmoid. We then find around the
center the local maxima and in that way define the boundary of the
plasmoid. Measuring in such a way the distance between the center of the
plasmoid and the local maxima, we can define the boundary in terms of
this distance.
%% After this point, we define the contours from the value of
%% the local maxima which is closer to the center.
Finally, as a validation check, we calculate the convex hull of these
contours to verify that the center of the plasmoid is indeed surrounded
by the configuration we have found.

\begin{figure*}
  \begin{center}
    \includegraphics[width=0.75\textwidth]{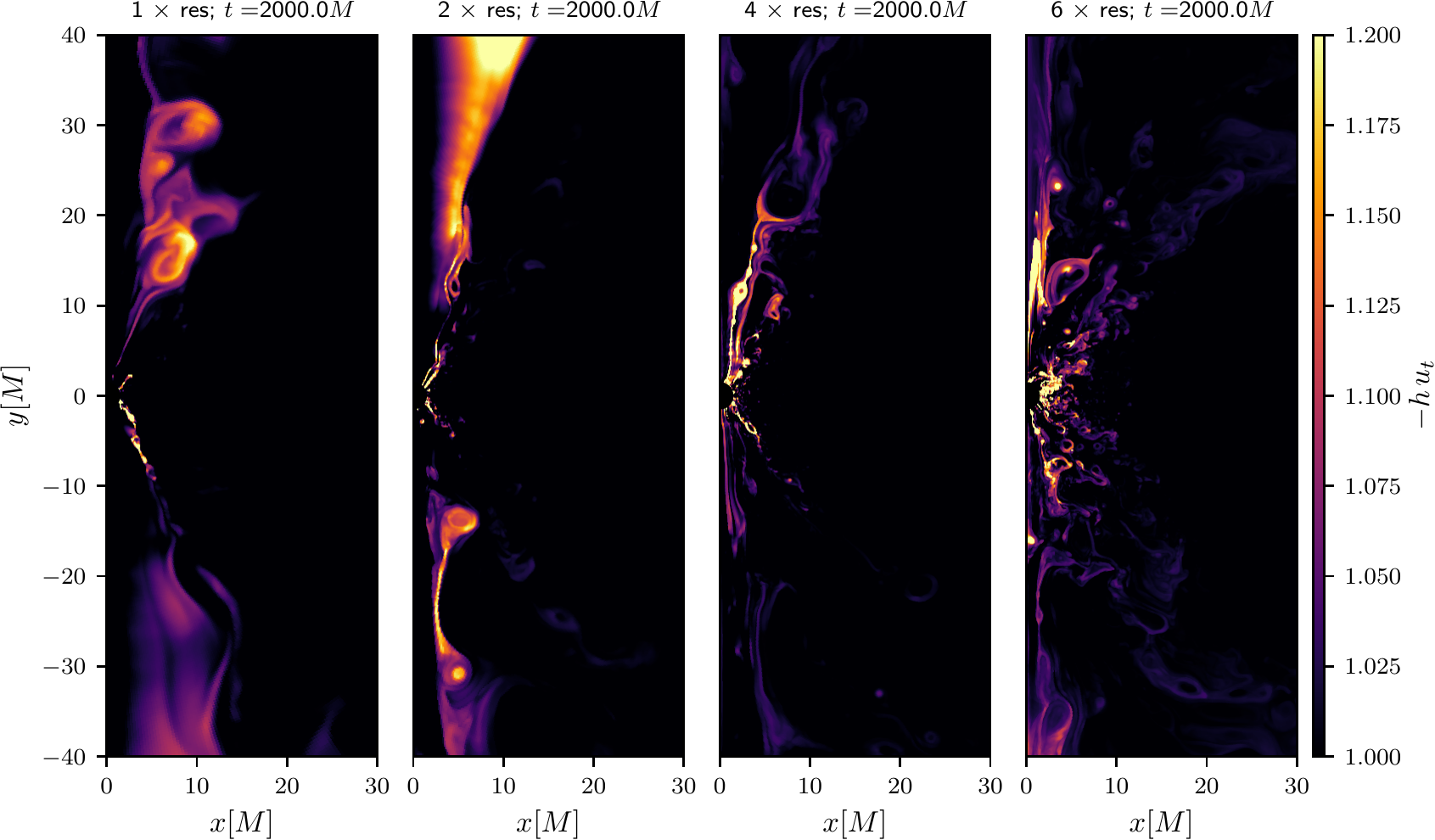}
    \includegraphics[width=0.75\textwidth]{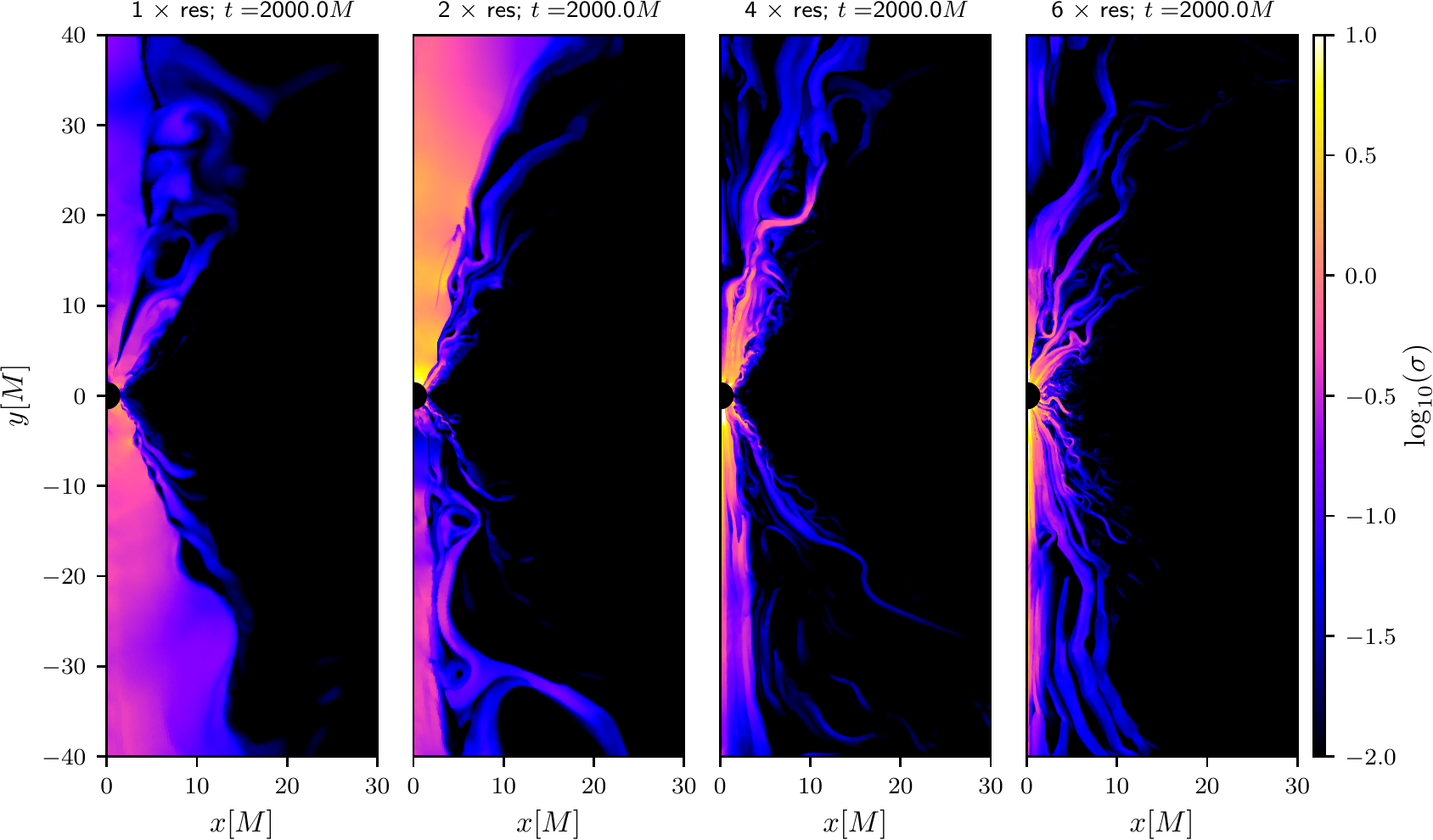}
  \end{center}
  \caption{The Bernoulli constant $-hu_t$ and the magnetization
    $\sigma$ for \texttt{\small{Model D}} at time $t=2000 \, M$ for the
    four different resolutions stated in Table \ref{table}.}
    \label{fig:res_hut}
\end{figure*}

In the three panels of Fig. \ref{fig:hut_8} we have presented the
tracking of a single plasmoid in terms of the Bernoulli constant, of the
toroidal magnetic field, and of the magnetization. In all snapshots
except one, the boundary is found after applying the Canny-edge detector
algorithm. However, in the snapshot with time $t= 5030\, M$, the plasmoid
boundary is not easily identified, which make it difficult to define a
closed boundary from the edges; in this case, the calculation of convex
hull was used. As a result, that for this specific snapshot, the shape of
the boundary of the plasmoid resembles a polygon.

%-----------------------------------------------------------------------
\section{On the sustainability of MRI turbulence in the torus}
\label{appen}
%-----------------------------------------------------------------------

We here provide evidence that the MRI is properly resolved throughout the
simulations by focussing only in the runs with the multi-loop magnetic
field structure with the alternating polarity, \ie \texttt{\small{Model
    D}} at the three resolutions in Table \ref{table}. A well-resolved
MRI is essential, since it provides a sustained source of turbulence and
-- in turn -- a quasi-stationary accretion process with an accretion rate
that is roughly constant in time.

As customary in these cases, we evaluate the so-called ``quality factor''
in terms of the ration between the grid spacing in a given direction
$\Delta x_{\theta}$, (\eg the $\theta$-direction) and the wavelength of
the fastest growing MRI mode in that direction (\ie $\lambda_{\theta}$),
where both quantities are evaluated in the tetrad basis of the fluid
frame $e_{\mu}^{(\hat{\alpha})}$ (see \citealt{Takahashi:2008,
  Siegel2013, Porth2019}, for details)
\begin{align}
  Q_{\theta}:=\frac{\lambda_{\theta}}{\Delta x_{\theta}}\,,
\label{Qth}
\end{align}
where
\begin{equation}
\lambda_{\theta}:=\frac{2\pi}{\sqrt{(\rho h +b^2)}\Omega}
b^{\mu}e_{\mu}^{(\theta)}\,,
\end{equation}
$\Omega:=u^{\phi}/u^t$ is the angular velocity of the fluid and the
corresponding grid resolution is $\Delta x_{\theta}:=\Delta
x^{\mu}e_{\mu}^{(\theta)}$.

\begin{figure*}
  \begin{center}
    \includegraphics[width=0.66\columnwidth]{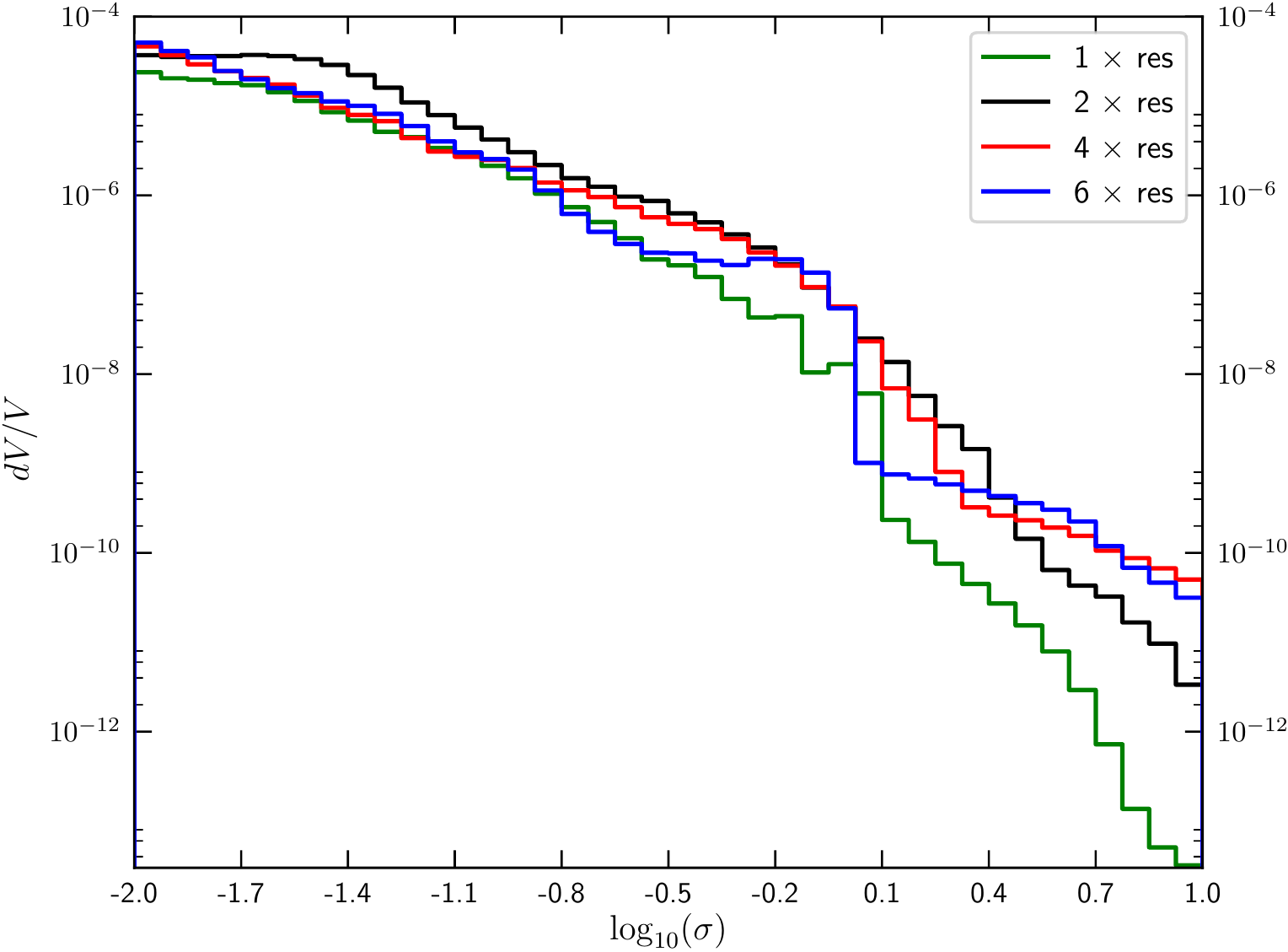}
    \includegraphics[width=0.66\columnwidth]{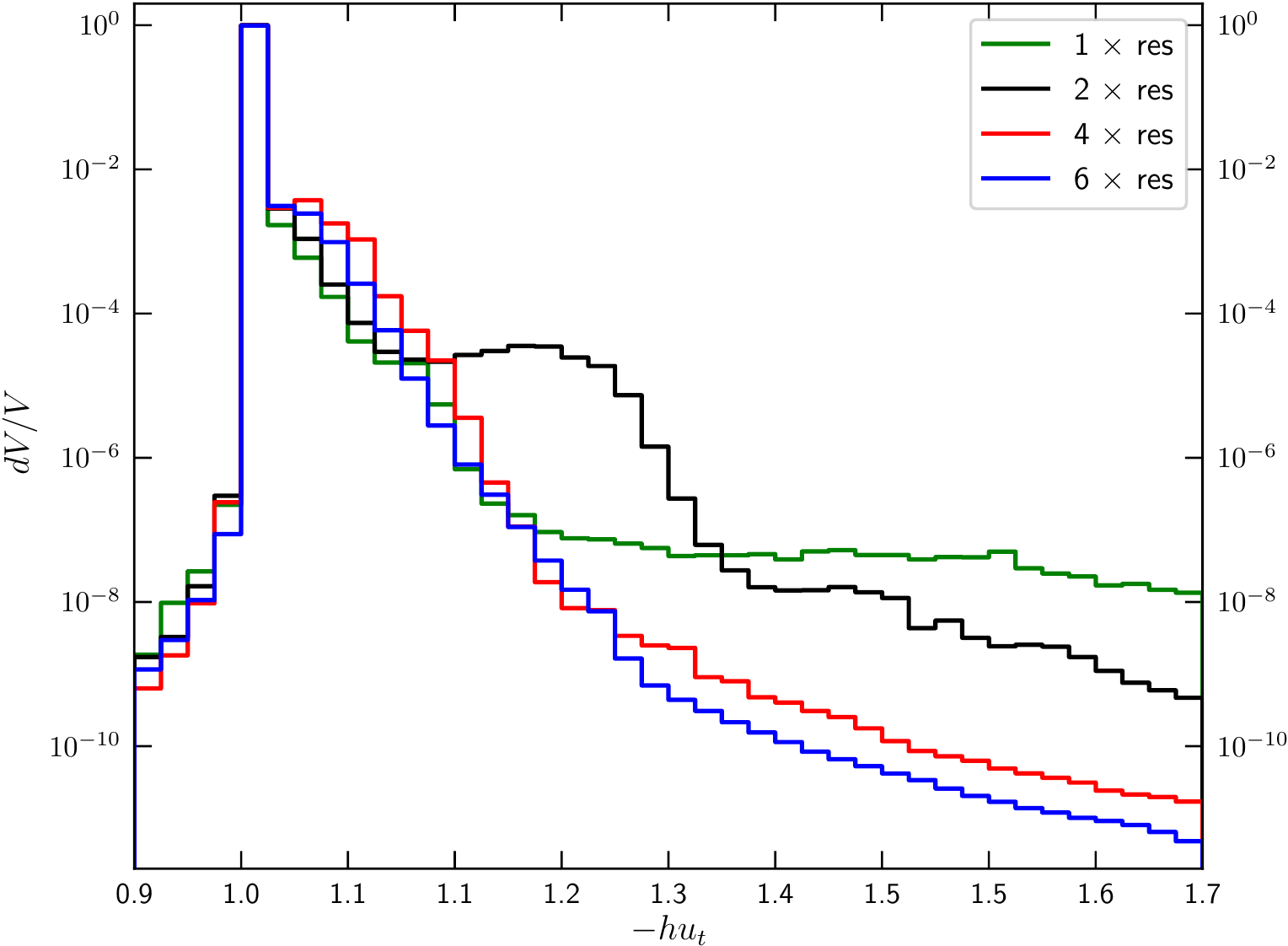}
    \includegraphics[width=0.66\columnwidth]{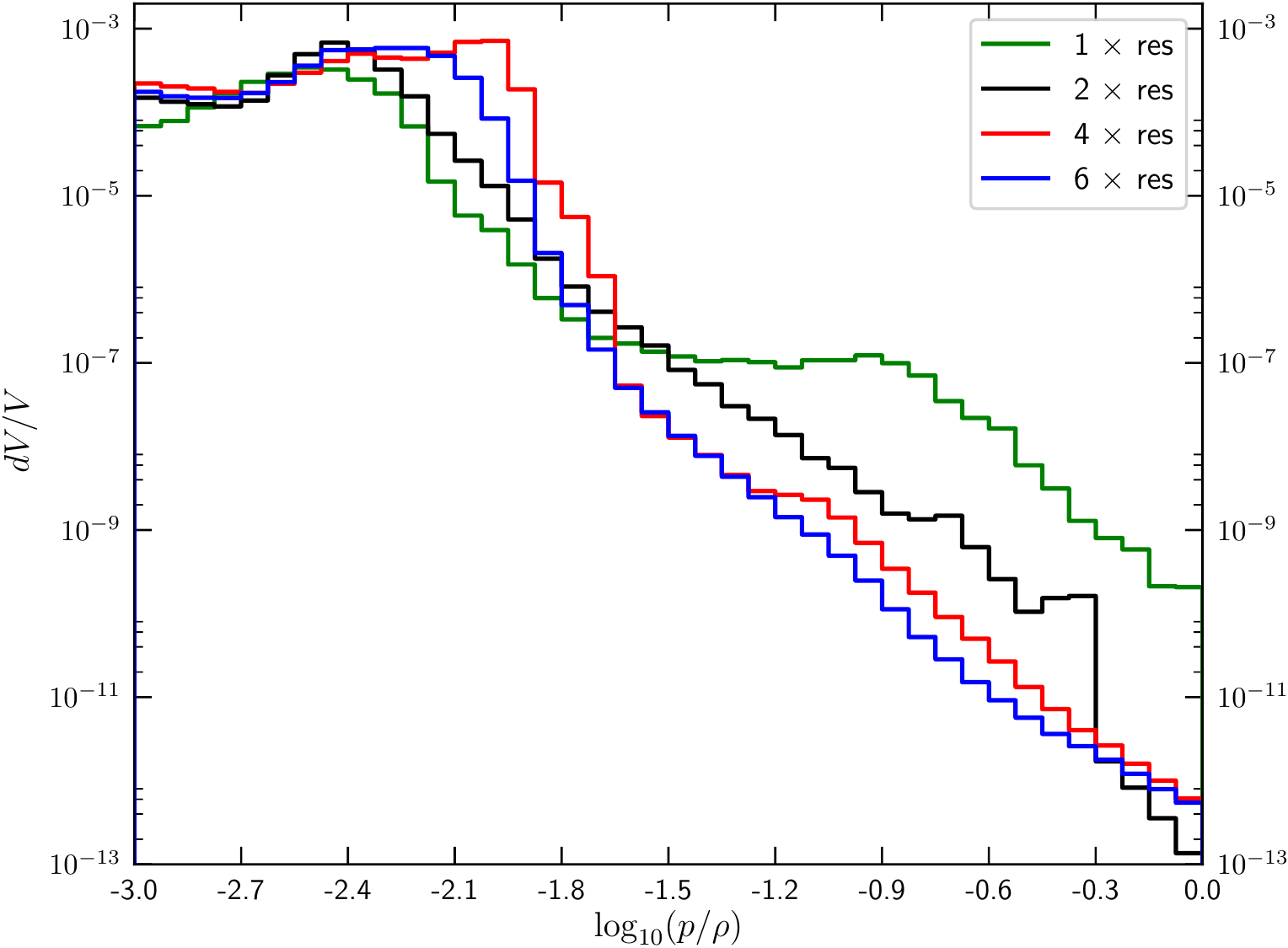}
  \end{center}
  \caption{Volume fractions of the magnetization parameter $\sigma$,
    Bernoulli constant $-hu_t$, and temperature proxy $p/\rho$ for
    \texttt{\small{Model D}} at the four different resolutions and over
    the time window $t=2000 - 3000\, M$.}
    \label{fig:hist_sigma}
\end{figure*}

The spatial distributions of the quality factor for three different
resolutions $1\times$, $2\times$, and $4\times$ the base resolution are
shown in the various panels of Fig. \ref{fig:Q} at time $t\approx 3000\,
M$. The left and middle panels, which refer to low and middle
resolutions, show that in these cases and at the time considered the
turbulence has started to decay and indeed the MRI is under-resolved in the
torus, as shown by the fact that it appears as mostly filled in
purple. On the other hand, the high-resolution run (right panel in
Fig. \ref{fig:Q}) shows that the MRI is in this case well resolved,
matching the expectation that with at least six cells covering
$\lambda_{\theta}$, the MRI is effectively resolved
\citep{Sano2004}. Furthermore, at this resolution, the accretion rate
maintains a rather constant value till time $t=12000\,M$, which is much
longer than the typical timescale investigated here (\ie $t=5 \, \times
\, 10^3\,M$).

\begin{figure*}
  \begin{center}
    \includegraphics[width=0.66\columnwidth]{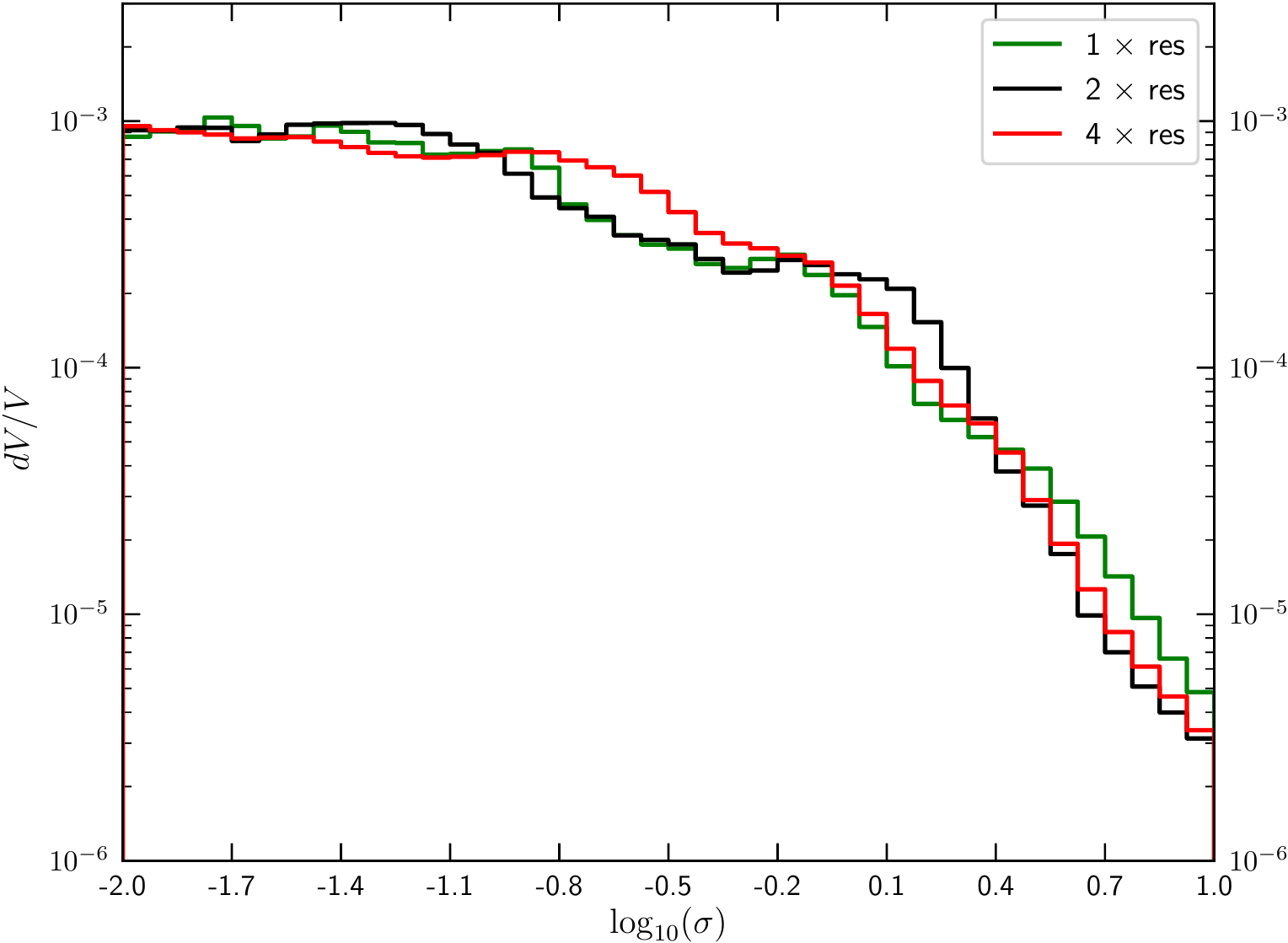}
    \includegraphics[width=0.66\columnwidth]{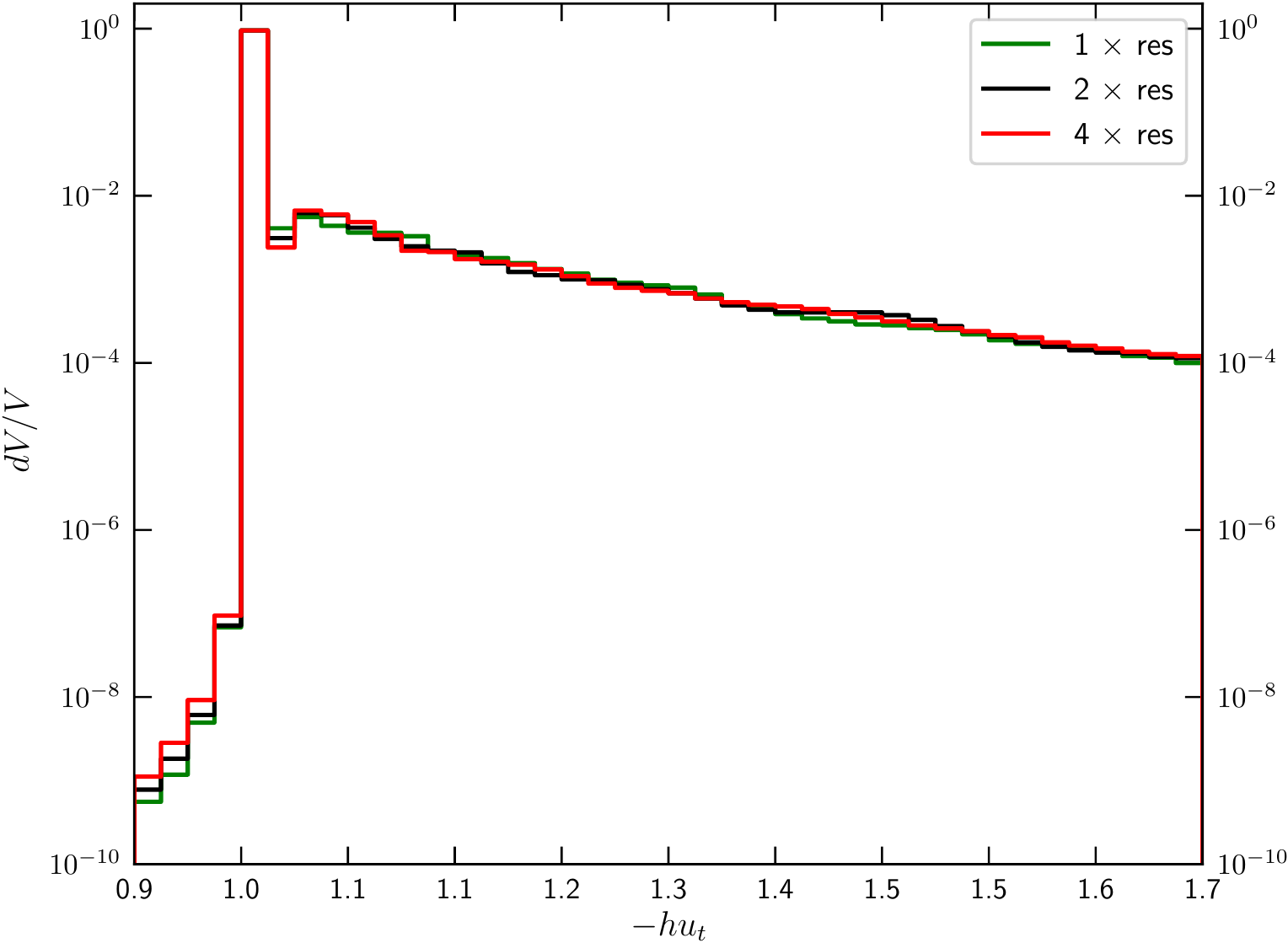}
    \includegraphics[width=0.66\columnwidth]{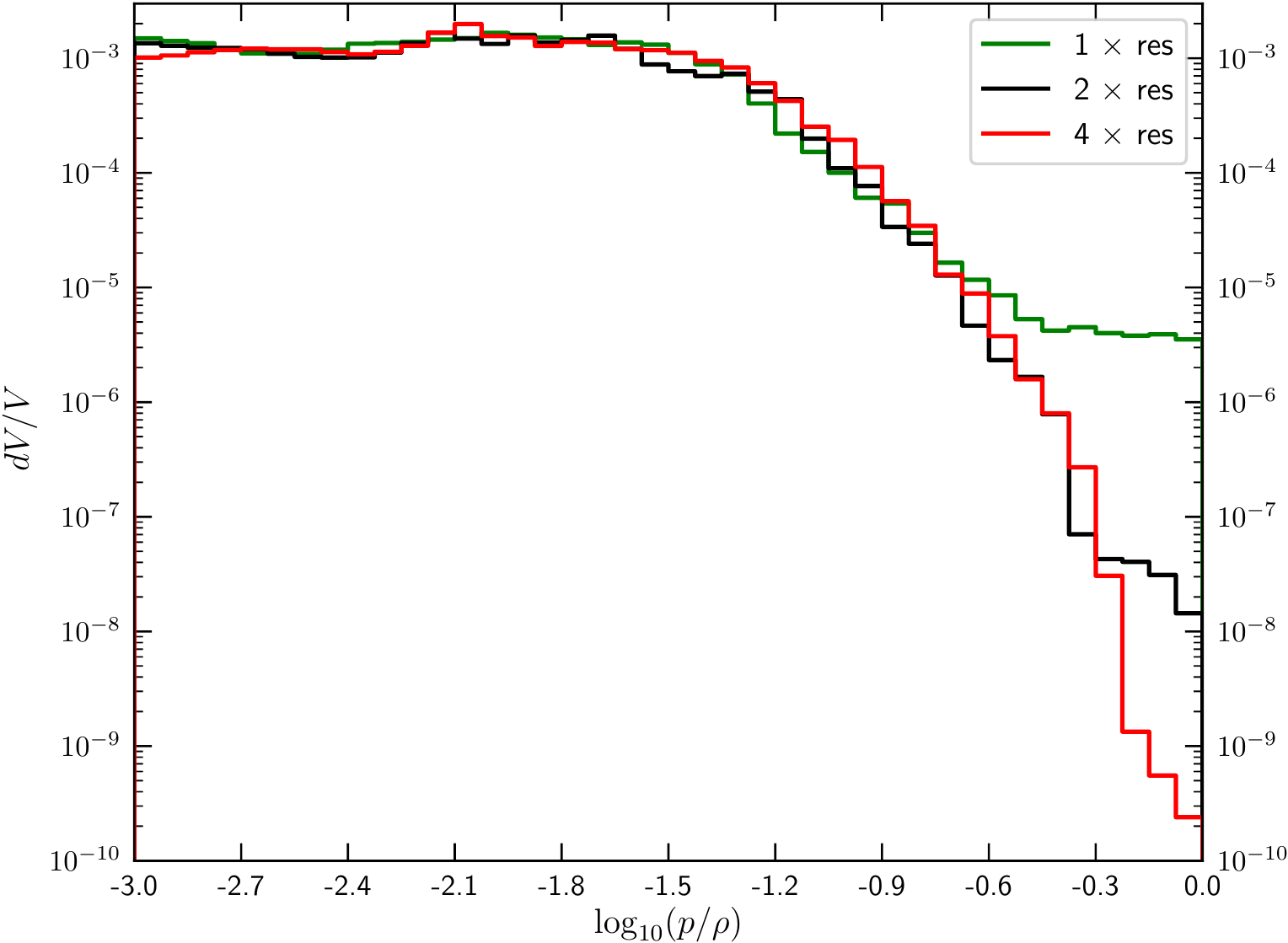}
  \end{center}
  \caption{Volume fractions of the magnetization parameter $\sigma$,
    Bernoulli constant $-hu_t$, and temperature proxy $p/\rho$ for
    \texttt{\small{Model A}} at three different resolutions ($1\times$, 
    $2\times$, and $4\times$ the base resolution) and over
    the time window $t=2000 - 3000\, M$.}
    \label{fig:hist_sigma_A}
\end{figure*}

\section{Dependence on resolution and numerical resistivity}
\label{appen:res}

Although resistivity plays a very important role in astrophysical
scenarios in general and in accretion disks in particular, its precise
value and dependence on the properties of the plasma (\eg temperature and
density) is still poorly known \citep{Fleming2000, Harutyunyan2018}. In
this Appendix we discuss the dependence of our results on resolution,
since in our simulations -- which solve the equations of ideal-MHD -- the
resistivity is purely numerical. At the same time, we recall that
previous work has explored the effect of resistivity on the formation and
growth rate of plasmoids and at which specific values the ``plasmoid
regime'' -- \ie the regime where plasmoid growth is exponentially rapid
-- takes place \citep{Ripperda2019b}. Furthermore, it has been argued
that in ideal-MHD, numerical resistivity yields results comparable with
the analytic expectations \citep{Obergaulinger:2009}.

It is important to note that plasmoids in \texttt{\small{Model D}}
 exist at all resolutions 
that we have tested, which are relatively high for GRMHD simulations
\citep{Porth2019, Akiyama2019_L5}. 

In the lower panels of Fig. \ref{fig:res_hut} we plot the magnetization
parameter $\sigma$ for the all four resolution for \texttt{\small{Model
    D}} at a time $t=2000 \, M$. As expected, from the previous
discussion, the first two runs show a much different distribution of
magnetization. In the two high resolution runs the magnetization has a
very similar structure. This structure seems to be affected by the
plasmoid production and the actual size of the plasmoids. At later times
a difference in magnetization is also evident for the high resolution
runs. However, due to the turbulent nature of the processes leading to
reconnection and thus to the production of plasmoids, a single snapshot
at a given time cannot fully illustrate and capture any differences in
magnetization due to resolution.

For a quantitative comparison of the whole evolution of
magnetization and the effective heating of the plasmoids, we have 
computed the distribution functions of the volume fraction 
$dV/V$ as a function of those quantities that are more sensitive to 
changes in resolution, namely, $\sigma, -hu_t$ and
$p/\rho$. Figure \ref{fig:hist_sigma} reports these distributions 
for \texttt{\small{Model D}}, over the timeframe 
$t=2000 \,-\, 3000\, M$, and shows that for the two runs with
lower resolution, the distribution of the volume fractions 
 are very different for all quantities. This is not the case
 for the two runs with higher resolution,
where both results agree well in highly magnetized regions.
This shows that the resolutions used are sufficient, not only 
for resolving the MRI, but also for a robust description of the 
plasmoid production and evolution. Similar distributions are shown 
in Fig. \ref{fig:hist_sigma_A} for \texttt{\small{Model A}},
where it is possible to appreciate that the plasma is well 
described already with the base resolution and that tori 
with single nested loops generically produce flows with 
larger magnetization, stronger outflows and higher 
internal energies.

\label{lastpage}
\end{document}